\documentclass[letterpaper,twocolumn,aps,prb,floatfix,
  citeautoscript,nobibnotes]{revtex4-1} 
\usepackage{epsfig,epstopdf}
\usepackage{graphicx}
\usepackage{verbatim}
\usepackage{amsfonts,amsmath,amssymb}
\usepackage{latexsym}
\usepackage{bm}
\usepackage{color}

\newcommand*\chem[1]{\ensuremath{\mathrm{#1}}}
\newcommand{\DTO}{\chem{Dy_2Ti_2O_7}}
\newcommand{\HTO}{\chem{Ho_2Ti_2O_7}}
\newcommand{\TTO}{\chem{Tb_2Ti_2O_7}}
\newcommand{\NIrO}{\chem{Nd_2Ir_2O_7}}
\newcommand{\HoIrO}{\chem{Ho_2Ir_2O_7}}
\newcommand{\RTO}{\chem{R_2M_2O_7}}
\newcommand{\OO}{\chem{O^{-2}}}

\begin{document}

\title{Numerical Simulations of the Magnetodielectric response in Ising Pyrochlores}

\author{T. Vignau Costa}
\affiliation{Instituto de F\'{\i}sica de L\'{\i}quidos y Sistemas Biol\'ogicos (IFLYSIB), UNLP-CONICET, B1900BTE La Plata, Buenos Aires, Argentina}
\affiliation{Departamento de F\'{\i}sica, Facultad de Ciencias Exactas, Universidad Nacional de La Plata, c.c. 16, suc. 4, B1900AJL La Plata, Argentina}

\author{S. A. Grigera}
\affiliation{Instituto de F\'{\i}sica de L\'{\i}quidos y Sistemas Biol\'ogicos (IFLYSIB), UNLP-CONICET, B1900BTE La Plata, Buenos Aires, Argentina}
\affiliation{Departamento de F\'{\i}sica, Facultad de Ciencias Exactas, Universidad Nacional de La Plata, c.c. 16, suc. 4, B1900AJL La Plata, Argentina}

\author{R. A. Borzi}
\affiliation{Instituto de F\'{\i}sica de L\'{\i}quidos y Sistemas Biol\'ogicos (IFLYSIB), UNLP-CONICET, B1900BTE La Plata, Buenos Aires, Argentina}
\affiliation{Departamento de F\'{\i}sica, Facultad de Ciencias Exactas, Universidad Nacional de La Plata, c.c. 16, suc. 4, B1900AJL La Plata, Argentina}

\email[Corresponding author: ]{borzi@fisica.unlp.edu.ar}

\begin{abstract}

In this paper, we examine the magnetoelectric response of Ising pyrochlores, focusing on both the ordered antiferromagnetic state and the frustrated ferromagnetic case known as ``spin-ice". We employ a model which accounts for magnetoelastic effects by considering the interplay between oxygen distortions and superexchange magnetic interactions within pyrochlores. This, together with numerical simulations, provides a tool to make quantitative comparisons with experiments. Our main target is then to see how to extract relevant information from this simple model, and to explore its limitations.
We obtain a direct estimation of quantities such as the electric dipole moment, the central oxygen displacement and the effective magnetoelastic energy for the canonical spin-ice material \DTO. We also inquire about the possibility of using the electric dipole carried by magnetic monopoles to obtain a direct measure of their density. In each studied scenario the correlations between monopoles, induced by their number or by the magnetic background,renders these findings less straightforward than initially anticipated. Furthermore, the coupling between electrical and magnetic degrees of freedom provides additional tools to investigate magnetic order in these systems.  As an example of this we discuss the phase diagram of the antiferromagnetic pyrochlore under applied magnetic field along the [111] direction.  We find an instance where the phase stability at nonzero temperatures is not dictated by the energy associated with different ground states but (akin to the phenomenon of {\em order-by-disorder}) is instead determined by their accessibility to thermal fluctuations.

\end{abstract}

\maketitle

\section{Introduction}

With their exponentially degenerate ground state manifold and exotic excitations, geometrically frustrated magnetic systems have been the focus of considerable attention in recent years~\cite{Diep2013frustrated,Lacroix2011introduction,Ramirez1994review,Moessner2006review}. The so called spin-ice materials, Ising ferromagnets with a pyrochlore structure, are among the best studied within this group~\cite{Bramwell01,udagawa2021spin}. 
Some members of this family can be grown as large single crystals~\cite{prabhakaran2011crystal}, while the relevant low temperature physics can be modeled by a classical and relatively simple Hamiltonian~\cite{melko2004monte,yavors2008dy,Borzi2016intermediate,henelius2016refrustration,samarakoon2020machine}; this has attracted the attention of a great number of experimentalists and theoreticians (see Ref.~\onlinecite{udagawa2021spin} and references therein). The low temperature ground state of pure spin-ice materials is exponentially degenerate, with the same entropy than that calculated by Pauling for water ice~\cite{ramirez1999zero}. In turn, this manifold can be thought as the magnetically neutral background where localized energy excitations analogous to magnetic charges --usually referred to as ``monopoles''-- move~\cite{Castelnovo,morris2009dirac}. There exist both single and double monopoles of each sign, the smaller magnetic charges being the lowest energy quasiparticles. Interestingly, the monopolar picture can be applied not only to spin-ice, but also to ``all-in/all-out'' (AIAO) Ising antiferromagnets~\cite{Guruciaga14}. Here the picture is reversed: the ground state corresponds a crystal of double monopoles with the structure of Zn-blende, the single monopoles remain the lowest energy excitations, and the neutral regions are now those with the highest energy. As it is the case in spin-ice~\cite{Castelnovo}, single monopoles can be stabilized in antiferromagnets by applying a magnetic field~\cite{Guruciaga16,pearce2022monopoledensity}. They can also be favored by distortions, both externally induced~\cite{pili_2019} or spontaneous ~\cite{jaubert2015holes,slobinsky2021monopole}.  

These magnetic excitations have also an effect on the elastic, and therefore the electrical, degrees of freedom. As it was theoretically demonstrated nearly ten years ago~\cite{Khomskii2012electric}, a combination of magnetic frustration and local asymmetry gives rise to a localized lattice distortion, with an associated electric dipole, for each single-charge monopolar excitation. On the other hand, double monopoles and neutral regions of the crystal lattice remain locally undeformed. There is some experimental evidence of the existence of the distortions accompanying monopoles~\cite{grams2014critical,jin2020experimental}. The estimation of their size (on the order of $0.1$ picometer) has been made indirectly, either through the electric dipolar energy needed to stabilize certain exotic monopolar phases in \TTO~\cite{Jaubert2015crystallog} (which, strictly speaking is not a spin ice material), or, in \DTO, by combining a magnetoelastic model with experiments on the dependence of the exchange constant with uniaxial deformation in ~\cite{slobinsky2021monopole}.

Due to the presence of the associated electric dipoles, it is to be expected that part of the magneto-dielectric response measured in spin-ice ~\cite{Katsufuji2004magnetocapacitance,Saito2005magnetodielectric,liu2013multiferroicity,lin2015experimental,yadav2019magnetodielectric} is related to magnetic charges. However, it is still an open question if magnetoelectricity due to monopoles is strong enough to be detected in standard macroscopic measurements, or how they can be used to extract information on their related magnetic properties. For example, using measurements on single crystals of the conducting \HoIrO\ combined with dipolar Monte Carlo simulations, it has recently been found that the isothermal magnetoresistance is highly sensitive to the monopole density~\cite{pearce2022monopoledensity}. This opens the question on whether the electric dipolar moment carried by single monopoles could be used as way to measure their density within a crystal of spin-ice or in a AIAO antiferromagnet.

In this work we address some of the points raised above. In order to do so we will review the Magnetoelastic Spin Ice (MeSI) model presented in Ref.~\onlinecite{slobinsky2021monopole} and discuss its use as a tool for the study of magnetoelectricity in Ising pyrochlores (Sec.~\ref{sec:Sys_and_Mod}). By means of this model we will extract quantitative information about the electric response of magnetic monopoles through the fluctuations of their dipolar electric moment, and will trace parallels with previous works studying the effect of these moments on phase stability~\cite{Jaubert2015crystallog} or their electric response~\cite{khomskii2021electric}. In Sec.~\ref{sec:AIAO} we will study using Monte Carlo simulations the magnetoelectricity of the AIAO antiferromagnet in zero magnetic field and with an external magnetic field applied in different directions. Sec.~\ref{sec:Ferromag} addresses the same problems in the case of spin-ice materials, taking profit of some published experimental results to  give quantitative estimates of the magnitude of the distortion and of the electric dipolar moment. Section~\ref{Sec:Discuss} considers these results, focusing first on the ability of magnetoelectricity to provide a direct measurement of the density of single magnetic monopoles (and of \textit{neutral} sites in AIAO antiferromagnets at low temperatures). We then provide a study, perhaps long overdue, on the phase diagram for the simplest possible AIAO model in an applied field, that will also contribute to the currently active field of antiferromagnetic iridates~\cite{opherden2017evolution,pearce2022monopoledensity}. After discussing some consistency checks for the assumptions made on the MeSI model, evaluating its main parameters, arguing about its limits, and discussing some surprising effects of monopole correlations, we summarize this work in Sec.~\ref{sec:Summ}.

\section{System and model}\label{sec:Sys_and_Mod}

\begin{figure}[htp]
    \includegraphics[width=0.60\columnwidth]{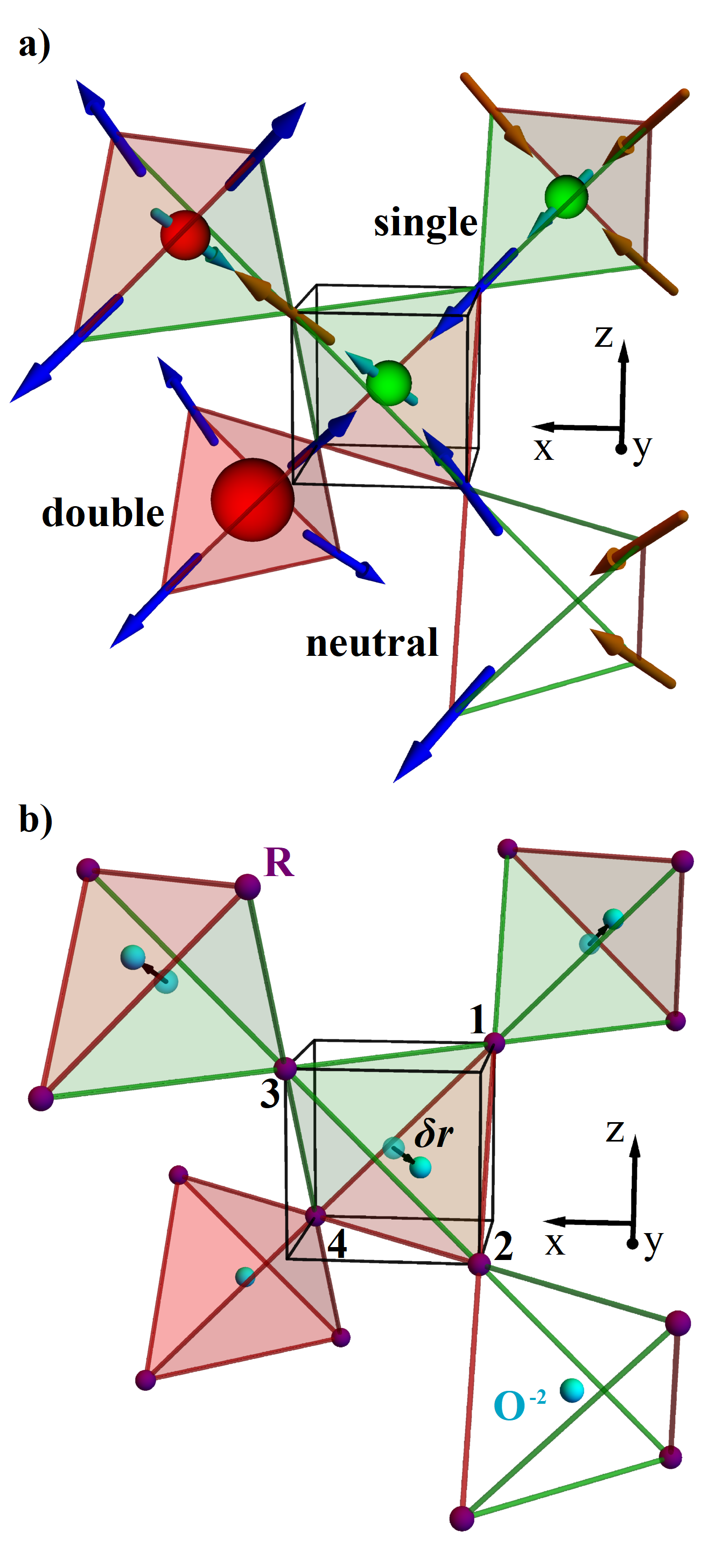}
    \caption{\textbf{Structure, magnetic monopoles and \OO-distortions.} \textbf{a)} Pyrochlore structure, with Ising spins in the shared vertices of up tetrahedra and down tetrahedra. The 3 in - 1 out configuration in the up tetrahedron (embedded in a cube) has associated a positive single monopole in its center (small green sphere); we also show single and double negative monopoles (small and big red spheres, respectively) and one neutral site (2 in - 2 out).  \textbf{b)} For single monopoles, the displacement $\delta \textbf{r}$ of the central oxygen ion (cyan sphere) along the cube diagonals decreases the exchange constants value along the three magnetic bonds it approaches (red lines and surfaces), and strengthens the other three (green lines and surfaces). No displacement occurs for neutral sites or double monopoles. Within the model, the rare earth ions (R) are assumed to be fixed.}
    \label{fig:Tetra}
\end{figure}

The pyrochlore lattice can be described as a cubic diamond lattice of corner sharing tetrahedra (Fig.~\ref{fig:Tetra}). Classical Ising magnetic moments, ${\boldsymbol \mu}_i =\mu {\mathbf S}_i  =\mu S_i \hat{\mathbf s}_i $, sit on the vertices of the tetrahedra with quantization directions $\hat {\mathbf s}_i$ along the local $\langle 111 \rangle$ directions. The pseudospins $S_i = \pm 1$ indicate if the magnetic moments point outwards ($+1$) or inwards ($-1$) of  ``up'' tetrahedra (embedded in a cube in Fig.~\ref{fig:Tetra}(a)). Magnetic charges occupy the centers of tetrahedra, labeled here using using Greek letters; their charge $Q_\beta$ is defined in direct proportion to the divergence of $\bm{S_i}$ across their surface. In this way, and as illustrated in Fig.~\ref{fig:Tetra}a), positive (negative) single monopoles belong to  3 in - 1 out (1 in - 3 out) tetrahedra, while positive (negative) double monopoles sit in all - in (all-out) ones, and neutral sites are related to 2 in - 2 out configurations.

\subsection{Magnetoelastic model}
The simplest magnetic Ising Hamiltonian on the pyrocholore lattice is the nearest neighbor model,

\begin{align}\label{eq:dsim}
{\mathcal H}_0^{\rm NN} \ = \ J_0 \sum_{\langle ij\rangle} S_i S_j \, ,
\end{align}\label{eq:nnsim0}
 
\noindent where $\langle \dots \rangle$ indicates that the sum is carried over nearest neighbours only. $J_0$ is an effective energy that takes into account possible contributions from superexchange and nearest neighbours magnetic dipolar interactions. A positive value of $J_0$ leads to frustration and a 2 in - 2 out locally neutral ground state that characterises spin ice materials~\cite{Bramwell01}. On the other hand the ground state for $J_0<0$ corresponds to the unfrustrated all-in/all-out antiferromagnet; in terms of its magnetic charge degrees of freedom it can be described as a  Zn-blende crystal of double monopoles. 

The superexchange interactions in \RTO\ pyrochlores are thought to be mediated by the oxygen \OO-ions sitting in the centre of the magnetic tetrahedra~\cite{Onoda2011exchange,Tomasello2018correlated,Sazonov2013,Jaubert2015crystallog}. We will assume for simplicity that the magnetic ions, corresponding to the rare earth atoms R, remain fixed in their pyrochlore lattice sites. Then, a displacement (see Fig.~\ref{fig:Tetra}b)) of the diamagnetic \OO-ions $\delta \textbf{r}_\beta$ will affect differently the effective exchange constant for each \textit{i-j} bond of the tetrahedron: $J_0 \xrightarrow{} J_{ij}(\delta \textbf{r}_\beta)$. Conversely, given a magnetic configuration for a tetrahedron, the \OO\ will displace its center of charge~\cite{Khomskii2012electric,Jaubert2015crystallog} so as to maximize the energy gain on satisfied bonds and minimize loses on unsatisfied ones. 

Following Ref.~\onlinecite{slobinsky2021monopole}, the simplest magnetoelastic Hamiltonian we can write is
\begin{align}\label{eq:HMagnetoElas}
{\mathcal H}_0^{\tiny \rm Me} = \sum_{\langle ij\rangle} J_{ij}(\delta \textbf{r}_\beta) S_i S_j + \sum_{\beta} \frac{1}{2} K \left(\frac{\delta r_\beta}{r_{nn}} \right)^2 \ ,
\end{align}\label{eq:nnsim}
\noindent where the \OO\ ions are modeled as sitting in a harmonic elastic potential with spring constant $K$ measured in kelvin, and  the different distortions $\delta \textbf{r}_\beta$ are approximated as uncorrelated from each other. We define the magnetoelastic constant, also measured in kelvin, as the change in the exchange energy of a bond when the intermediary \OO\ moves away from it; using (see Fig.~\ref{fig:Tetra}b)) the bond $2-4$ for concreteness, 
\begin{align}\label{eq:alpha}
\tilde{\alpha} \ \equiv \ r_{nn} \, \frac{\partial J_{24}}{\partial z}\Bigr|_{\delta \textbf{r}=0}  \ ,
\end{align}

\noindent where $r_{nn}$ is the distance between nearest neighbours.

Ref.~\onlinecite{slobinsky2021monopole} shows that, to first order in the distortions, the Hamiltonian in Eq.~\ref{eq:HMagnetoElas} leads to the so-called Magnetoelastic Spin Ice (MeSI) model. It consists of an effective spin-only part, ${\cal H}_0^{\rm eff}(\{S_i\})$, together with a modified elastic term that depends on both degrees of freedom, ${\cal H}^{\rm elas}(\{\delta \textbf{u}_\beta\}, \{S_i\})$:
\begin{align}\label{eq:HMesi}
{\cal H}_0^{{\tiny \rm Me}} \approx {\cal H}^{{\tiny \rm MeSI}} \ \equiv {\cal H}^{\rm eff}_0(\{S_i\}) + {\cal H}^{\rm elas}(\{\delta \textbf{r}_\beta\},\{S_i\}).
\end{align}
The magnetic Hamiltonian can be written as
\begin{align}\label{eq:HMag}
{\cal H}^{\rm eff}_0(\{S_i\}) \ \equiv \ \sum_{\beta} \Big( J_{ml} \prod_{i=1}^4 S_i + \frac{1}{2} J_0 \sum_{i \neq j = 1}^4 S_{i}S_{j} \Big) \ ,
\end{align}
\noindent where the index $\beta$ that sweeps up and down tetrahedra is left implicit in the pseudospin variables. 
The constant
\begin{align}\label{eq:Jml}
J_{ml} \ &\equiv \  \frac{3 \tilde{\alpha}^2}{K} 
\end{align}
represents a new effective magnetic energy scale in a \textit{four}-spin Hamiltonian (see Refs.~\onlinecite{jaubert2015holes} and~\onlinecite{Slobinsky2018charge}). It favors the creation of single monopoles against neutral sites or double charges; as it will be seen afterwards, the conditions implicit in our studies are such that this term will not play a major part in this work. 

In the pŕesence of an externally applied magnetic field, a Zeeman term should be added to the effective Hamiltonian:
\begin{align}\label{eq:Zee}
    {\mathcal H}^{{\tiny \rm Zeeman}} = - \sum_i \mu {\mathbf B} \cdot {\mathbf S}_i \, ,
\end{align}
with $\bm{B}$ the external magnetic field.
In order to simulate more accurately spin ice materials such as HTO or DTO, (Sec.~\ref{sec:Ferromag}) it is necessary to consider long-range dipolar interactions, 
\begin{align}\label{eq:dipm}
    {\mathcal H}^{{\tiny \rm Dip.M.}} = D\, r_{\textit{nn}}^3 \sum_{i>j}^{}{'} \left[ \frac{{\mathbf S}_i \cdot{\mathbf S}_j}{|{\bf{r}}_{ij}|^3}- \frac{3({\mathbf S}_i \cdot {\bf{r}}_{ij}) ({\mathbf S}_j \cdot {\bf{r}}_{ij}) }{|{\bf{r}}_{ij}|^5} \right] \, .
\end{align}
Here, the primed sum indicates the exclusion of the contribution to nearest neighbors interactions, which has already been taken into account in Eq.~\ref{eq:HMag}.

To better elucidate the elastic properties of the MeSI Hamiltonian, it is useful to re-write ${\cal H}^{\rm elas}$ in Eq.~\ref{eq:HMesi} as

\begin{align}\label{eq:Helas}
{\cal H}^{\rm elas}(\{\delta \textbf{r}_\beta\},\{S_i\}) \ & = \ \sum_{\beta} \frac{3}{2} J_{ml}^{-1} (\delta {\textbf{O}_\beta})^2 + {\rm const.}   \ , \\
\label{eq:deltaO}
\delta \textbf{O}_\beta  \ & \equiv \frac{\tilde{\alpha}}{ r_{nn}} (\delta{\textbf{r}}_\beta - \delta{\bm{r}}^{\rm eq}_\beta(\{S_i\}_\beta)) \ .
\end{align}
This term is quadratic on the distortion variable $\delta \bm{r}_\beta$. However, the feedback from the \textit{magnetic} configuration $\{S_i\}_\beta$ of the tetrahedron $\beta$ redefines the position of the elastic energy minimum for the \OO-ion: it is now displaced from the center of the tetrahedron by $\delta{\bm{r}}^{\rm eq}_\beta(\{S_i\}_\beta)$. 
For neutral or double monopoles, $\delta{\bm{r}}^{\rm eq}_\beta(\{S_i\}_\beta) = 0$: the $\OO$-ion remains in the center of its tetrahedron, and there is no local dipolar electric moment. However, for single monopoles~\cite{Khomskii2012electric} of any sign the energy balance dictates that there is a new equilibrium point given by~\cite{slobinsky2021monopole}
\begin{align}\label{eq:deltaRmin}
\delta{\bm{r}}^{\rm eq}_\beta(\{S_i\}_{\beta}) \ = \ \eta \frac{2r_{nn} \ J_{ml}}{\sqrt{3}\Tilde{\alpha}} \ \hat{d} (\{S_i\}_\beta) \ ,
\end{align}
where $\eta$ takes the value +1 (-1) for up (down) tetrahedra and
\begin{align}\label{eq:diagonals}
\hat{d} (\{S_i\}_\beta) \equiv \hat{d}_\beta =
\begin{cases}
\frac{1}{\sqrt{3}} (-1, -1, -1)\\
\frac{1}{\sqrt{3}} (-1, \ 1, \ 1)\\
\frac{1}{\sqrt{3}} (\ 1,\ 1, -1)\\
\frac{1}{\sqrt{3}} (\ 1, -1, \ 1) \ .
\end{cases}
\end{align}
The mean displacement direction of the \OO\ is along the direction of the \textit{minority} spin in a single magnetic charge, and towards the triangular phase of the tetrahedron with three ``in-in'' or ``out-out'' bonds (painted red in Fig.~\ref{fig:Tetra}). According to the MeSI model then, the magnitude of this distortion (which we will be able to estimate in a real material) is determined by the ratio $J_{ml} / \Tilde{\alpha}$, proportional to $\Tilde{\alpha} / K$. 

\subsection{Magnetoelectric properties}
An important message from the previous subsection is that a large magnetoelastic coupling and small restoring forces not only favor the existence of single monopoles but also the existence of larger average distortions with consequently bigger microscopic electric moments. 
In this paper we will consider $J_{ml} / \Tilde{\alpha}$ to be large enough for the electric phenomena associated with single monopoles to be detectable, but small enough to make electric dipolar interactions negligible (some consequences of these interactions have been studied in Ref.~\onlinecite{Jaubert2015crystallog}, and further developed within the MeSI model in~\onlinecite{Vignau23}). Furthermore, as opposed to Ref.~\onlinecite{slobinsky2021monopole}, here we will work within the weak limit given by $J_{ml}\ll J_0$.  Therefore, from this point onwards we will neglect the four-spin term in Eq.~\ref{eq:HMag}; this assumption will be checked for consistency for spin ice materials in Sec.~\ref{Sec:Discuss}. 

We now concentrate on the magnetoelectric properties of the \OO-ions involved in super-exchange, which manifest through their electric dipolar moments $\bm{p}_\beta$. We define the contribution to the electric dipolar moment associated to central \OO\ ions along the unit direction $\hat{e}$ as:
\begin{align}
\label{eq:p_q}
P^{\OO}_{\hat{e}} \ \equiv  \hat{e} \cdot \sum_\beta  \bm{p}_\beta \  = \ -2q\, \hat{e} \cdot \sum_\beta \ \delta \bm{r}_\beta \\
= \ {P}^{\rm up}_{\hat{e}} + P^{\rm do}_{\hat{e}} \notag \ ,
\end{align}
where $q$ is the elementary electric charge.  In the last line we have separated the contributions from \OO-ions in up and down tetrahedra.
Its equilibrium fluctuations are connected with the electric response of a sample of volume $V$ to a field, the static electric susceptibility, 
\begin{align}\label{eq:chi_P}
\chi^{\OO}_{\hat{e}} \ = \  \frac{1}{\varepsilon_0 Vk_BT} \ \big( \big\langle (P^{\OO}_{\hat{e}})^2 \big\rangle - \big\langle P^{\OO}_{\hat{e}} \big\rangle^2 \big) \ .
\end{align}
It is interesting to see how the MeSI Hamiltonian gives a physical interpretation to the procedure used  in Ref.~\onlinecite{Jaubert2015crystallog} to study some of the electrical properties of \TTO.  One can split the \OO-ion distortions into two terms, the first one corresponding to the displacement of the magnetoelastic minimum determined by the spin configuration,  
and a second  one corresponding to thermal (or eventually quantum) fluctuations that make it vibrate around it,
\begin{align}
    \delta \textbf{r}_\beta \ = \ \delta \textbf{r}^{\rm eq}_\beta(\{S_i\}_\beta) + \delta \textbf{v}^{\rm th}_\beta \ .
\end{align}
The average electric polarisation fluctuations (Eq.~\ref{eq:chi_P}) have no crossed term, leaving two contributions to the electric susceptibility. 
\begin{align}
\label{eq:chi_O-2}
\chi^{\OO}_{\hat{e}}(\{\delta \textbf{r}_\beta \}) \ =& \  \chi^{\rm mon}_{\hat{e}}(\{\delta \textbf{r}^{\rm eq}_\beta \}) \ + \ \chi^{\rm th}_{\hat{e}}(\{\delta \textbf{v}^{\rm th}_\beta(T) \})
\end{align}
\noindent The first one corresponds to the fluctuations of an electric dipolar moment of constant magnitude  $-2q \ \delta \bm{r}^{\rm eq}_\beta$ related to single monopoles, with \OO~ions precisely \textit{at} the new minimum of magnetoelastic energy. The second contribution is thermally or quantum mechanically activated and concerns all tetrahedra, independently of its topological charge. It depends on the fluctuations of the \OO~\textit{around} the minimum, $\langle (\delta v^{\rm th}_\beta)^2 \rangle$; through equipartition, it would be proportional to $T$ in a classical context. We will neglect it in our study of the contribution of monopoles to the electrical activity of pyrochlores, considering that we generally are in a temperature regime quite below the Debye temperature of the materials.

Using Monte Carlo simulations we will measure this susceptibility along $\hat{e}$ through its average thermal fluctuations: 
\begin{align}\label{eq:chi_mon}
\chi^{\rm mon}_{\hat{e}} \ \equiv \ D_q \frac{3}{T N_q} \ \big( \ \big\langle ( \sum_\beta \hat{d}_\beta \cdot \hat{e} )^2 \big\rangle - \big\langle \sum_\beta \hat{d}_\beta \cdot \hat{e}  \big\rangle^2 \ \big) \ .
\end{align}
Here,
\begin{align}\label{eq:De}
D_q \equiv \frac{\delta_q p_q^2}{3\varepsilon_0 k_B},
\end{align}
is measured in kelvin, $\delta_q=N_q/V$ is the number density of central \OO-ions, $\varepsilon_0$ the electric permittivity of vacuum, and $k_B$ Boltzmann constant. Note that the measurement of the electric susceptibility can give us access to the microscopic dipolar electric moment; $p_q$ in turn can reflect the materials magnetoelastic properties:
\begin{align}
    p_q = 2q \delta r^{\rm eq}_\beta = \frac{4q \ r_{nn}}{\sqrt{3}} \frac{J_{ml}}{\hat{\alpha}}. 
\end{align}
For brevity, we will sometimes note $\chi^{\rm mon}_{\hat{e}}$ using a non-unity vector (e.g., we note $\chi^{\rm mon}_{[111]}$). 

\subsection{Single tetrahedron approximation (no applied field)}

As it was discussed, within the MeSI model the only contribution to the electric susceptibility at low temperature is associated with single monopoles. It will be useful to compute this susceptibility for the case of a single tetrahedron where the probability of having a single monopole is given by the average density of single monopoles per tetrahedron in the pyrochlore lattice, $\rho_{\rm s}$. The electric susceptibility from Eq.~\ref{eq:chi_mon} is now 
\begin{align}
\chi^{\rm 1tet}_{\hat{e}} \ \equiv \  3\rho_{\rm s}(T) \frac{D_q}{T} \ \sum_\beta \big( \ \big\langle ( \hat{d}_\beta \cdot \hat{e} )^2 \big\rangle - \big\langle \hat{d}_\beta \cdot \hat{e}  \big\rangle^2 \ \big) \ .
\end{align}
If there is no symmetry breaking field applied, all spin configurations associated with single monopoles will be equally probable, leading to $\langle \hat{d}_\beta \rangle = 0$: only the quadratic fluctuations are then important. It can be shown that for the isotropic case the average, and thus the susceptibility, is independent of $\hat{e}$:
\begin{align}\label{eq:chi_one_mon}
\chi^{\rm 1tet}_{\hat{e}}(\bm{B}=0) \ = \ D_q \ \frac{\rho_{\rm s}(T)}{T}.
\end{align}
This expression should describe the electric response of a pyrochlore lattice where the single monopole magnetic configurations in different tetrahedra are independent of each other. In this case, the electric susceptibility can provide an indirect measure of the monopole density.


\subsection{Simulation details}

It is remarkable that within the framework we have discussed it is not necessary to use a complete magnetoelastic Hamiltonian in order to calculate the monopole contribution to the electric properties of spin ice like materials. Instead, we can restrict ourselves to the magnetic part of the MeSI Hamiltonian (${\cal H}^{\rm eff}_0(\{S\})$ in the simplest case, to which other pure magnetic terms like Eqs.~\ref{eq:Zee} or~\ref{eq:dipm} can be added). This is a great advantage from the viewpoint of computational physics.

Here, we performed Monte Carlo simulations with the Metropolis algorithm and single-spin-flip dynamics. In order to implement Eq.~\eqref{eq:dipm} in the algorithm, we used Ewald summations to take into account long-range interactions. The conventional unit cell of the pyrochlore lattice (Fig.~\ref{fig:Tetra}(a)) consists of $16$ spins, and we simulated cubic systems of $L^3$ cells with periodic boundary conditions. As an example, for a system with $L=4$ and dipolar interactions we took $8\times 10^5$ Monte Carlo steps for equilibration, and then $2\times 10^5$ steps were used to calculate the averages at each value of temperature and applied magnetic field. In turn, the results were averaged over $10$ independent runs.


\section{All-in/all-out Ising antiferromagnets in the pyrochlore lattice ($J_0<0$)}\label{sec:AIAO}
We start by studying the magnetoelastic properties of the MeSI model with antiferromagnetic nearest neighbours interactions, $\mathcal{H} = \mathcal{H}^{\rm eff}_0 + \mathcal{H}^{\rm Zeeman}$. Along this section $J_0<0$, and ---as for the rest of the paper--- we assume $J_{ml} \ll |J_0|$.

\subsection{Zero magnetic field}

\begin{figure}[htp]
    \includegraphics[width=0.80\columnwidth]{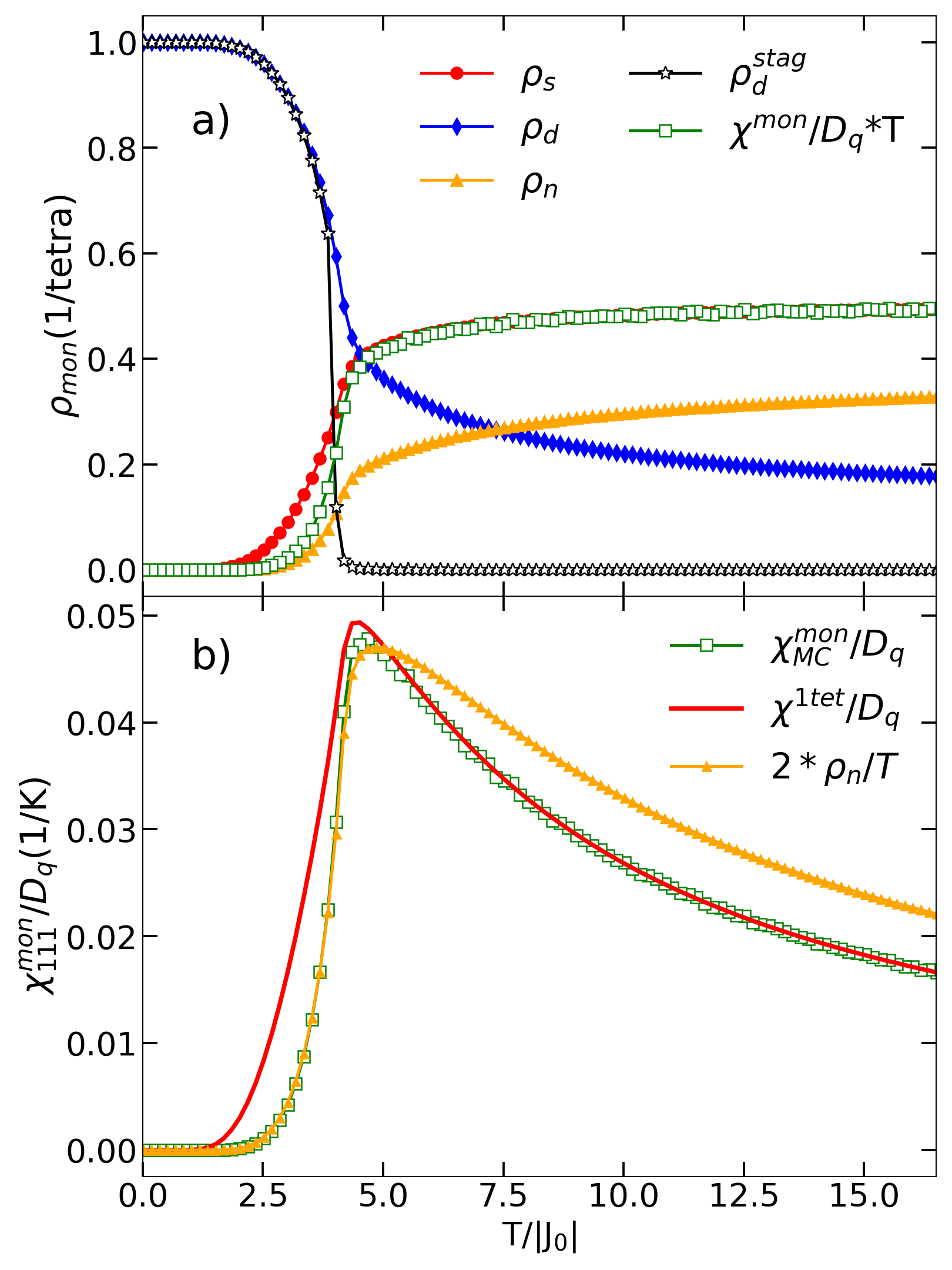}
    \caption{\textbf{Zero field magnetoelastic behaviour for the AIAO antiferromagnet from Monte Carlo simulations ($L=4$).} \textbf{a)} Density of magnetic charges as a function of temperature for single monopoles ($\rho_{\rm s}$), double monopoles ($\rho_{\rm d}$), and neutral sites $\rho_{\rm n}$ for $\mu B=0$. We also include the order parameter for the Zn-blende crystal of double monopoles, $\rho_{d}^{\rm stag}$, and an estimate for $\rho_{\rm s}$ taken from the electric response in Panel b), $\chi^{\rm mon}/D_q*T$. This estimation is excellent above the ordering temperature of the antiferromagnet, $T_C/J_0 \approx 4.2$; below $T_C$, the electric response is accounted for by the density of neutral sites (see Panel b)); \textbf{b)} Electrical susceptibility due to monopoles, $\chi_{\rm MC}^{\rm mon}$, (open symbols) calculated from Monte Carlo simulations compared with the single tetrahedron approximation $\chi^{\rm 1tet}$ based on the single monopole density (red line), and  with a similar estimate based on the density of neutral sites (orange symbols).  See text for details. } 
    \label{fig:AFB=0}
\end{figure}

We begin the study with the case of no applied external magnetic field.  Fig.~\ref{fig:AFB=0}a) shows the density of double monopoles (blue curve) per tetrahedron as a function of temperature and $B=0$.  As expected, at temperatures below  $T/J_0 \approx 5$ the density is compatible with the formation of a crystal of double charges. There are two possible antiferromagnetic domains; as we will see later, it will be useful to identify them by the sign of the  magnetic charge in the up tetrahedra. For the perfect crystal, a positive ($+2Q$) charge marks the all-in/all-out domain type, while a negative $-2Q$ magnetic charge identifies all-out/all-in (AOAI) domains. The formation of a crystal is confirmed by the order parameter associated to this phase. We define the staggered charge density of double monopoles $\rho_{\rm d}^{\rm stag}$ as the modulus of the total charge due to double monopoles in up tetrahedra per unit charge. We can see in Fig.~\ref{fig:AFB=0}a) how $\rho_{\rm d}^{\rm stag}$ raises from very near zero for temperatures below $T_C/|J_0| \approx 4.2$. 

We have mentioned that there is no intrinsic electric activity associated with double charges; however, there should be an electrical response from the crystal's lowest energy excitations: the single monopoles. Their density is measured by the red curve in Fig.~\ref{fig:AFB=0}a). Fig.~\ref{fig:AFB=0}b) shows that the electric susceptibility due to monopoles calculated using Eq.~\ref{eq:chi_mon} along $\hat{e}\parallel$[111] increases in a Curie-law fashion for decreasing temperature (green curve). It peaks near $T_C/|J_0|$, where it gets the best compromise between a relatively big density of single monopoles and minimum thermal disorder. At lower $T$ it drops suddenly (faster even than the density $\rho_{\rm s}$, as we will see below) as the Zn-blende structure of double monopoles becomes less defective. 

It is interesting that the single tetrahedron approximation, Eq.~\ref{eq:chi_one_mon}, drawn in red in Fig.~\ref{fig:AFB=0}b) reproduces the true susceptibility above $T_C$; however, $\chi^{\rm 1tet}$ overestimates it below the transition. Correspondingly, if we calculate the monopole density directly from the electric response assuming $\chi^{\rm 1tet}/D_q =  \rho_{\rm s}(T)/T \approx \chi^{\rm mon}/D_q$ we obtain the green curve in Fig.~\ref{fig:AFB=0}a). The approximation follows the behavior of the true $\rho_{\rm s}(T)$ at high $T$ really closely (they never differ by more than 5\%, value taken at $T_C$), but it is rather poor below it. A quick look at the density of the energetically more expensive neutral sites (orange symbols in Fig.~\ref{fig:AFB=0}b)) shows that the decreasing trend of $\chi^{\rm mon}/T$ as $T\to 0$ resembles more the behavior of $\rho_{\rm n}(T)$ than $\rho_{\rm s}(T)$.

While the difference between both susceptibility curves in Fig.~\ref{fig:AFB=0}b) is obviously due to correlation effects, it is interesting to discuss why their departure becomes noticeable below the crystallization temperature. The lowest energy excitations for the perfect crystal of double charges involves the flipping of a single spin, to produce two single monopoles linked by this \textit{minority} spin (see the central and upper left tetrahedron in Fig.~\ref{fig:Tetra}). Although two new single monopoles are created, and with them two new electric dipolar moments, it is easy to see that there is no associated net electric moment fluctuation. In other words: $P=0$ for the perfect crystal, since there are no dipoles; and $P=0$ after the fluctuation since the new dipolar moments cancel each other, in a fashion that recalls the discussion on the reduced electric response in a crystal of single monopoles in Ref.~\onlinecite{khomskii2021electric}. This reasoning explains the origin of the correlation, and the fact that the curve for independent tetrahedra (proportional to $\rho_{s}(T)$) overestimates the electric susceptibility, and underestimates the true value of $\rho_s$.

Given the constraints imposed by the construction rules of a double monopole crystal, it follows that a non-zero $\chi^{\rm mon}$ for $T<T_C$ necessarily implies the existence of other type of excitations. The flip of a second spin, one of the majority spins linking a single with a double monopole (see the double monopole on the lower left corner in Fig.~\ref{fig:Tetra}), separates and decorrelates the two original single monopoles. This second flip results in a neutral site between two single monopoles and in a non-zero total electric moment. Hence, the density of neutral sites results directly proportional (with a factor of $2$) to the density of \textit{electrically active} (i.e., decorrelated) single monopoles,
\begin{align}
    \chi^{\rm mon} \ = \ D_q \frac{2\rho_{\rm n}}{T} \ , \ T<T_C \ .
    \label{eq:neut_vs_chi}
\end{align}

Fig.~\ref{fig:AFB=0}b) shows that $2\rho_{\rm n}/T$ (orange triangles) is an excellent approximation for $\chi^{\rm mon}/(D_qT)$ (green open squares) below the crystallization temperature. Note also that below $T_C$, the difference $\rho_{\rm s}-2\rho_{\rm n}$ can be interpreted as the number of coupled, electrically inactive single monopoles.

\subsection{Magnetic field $\textbf{B} \parallel$ [111] }

\begin{figure}[htp]
    \includegraphics[width=0.8\columnwidth]{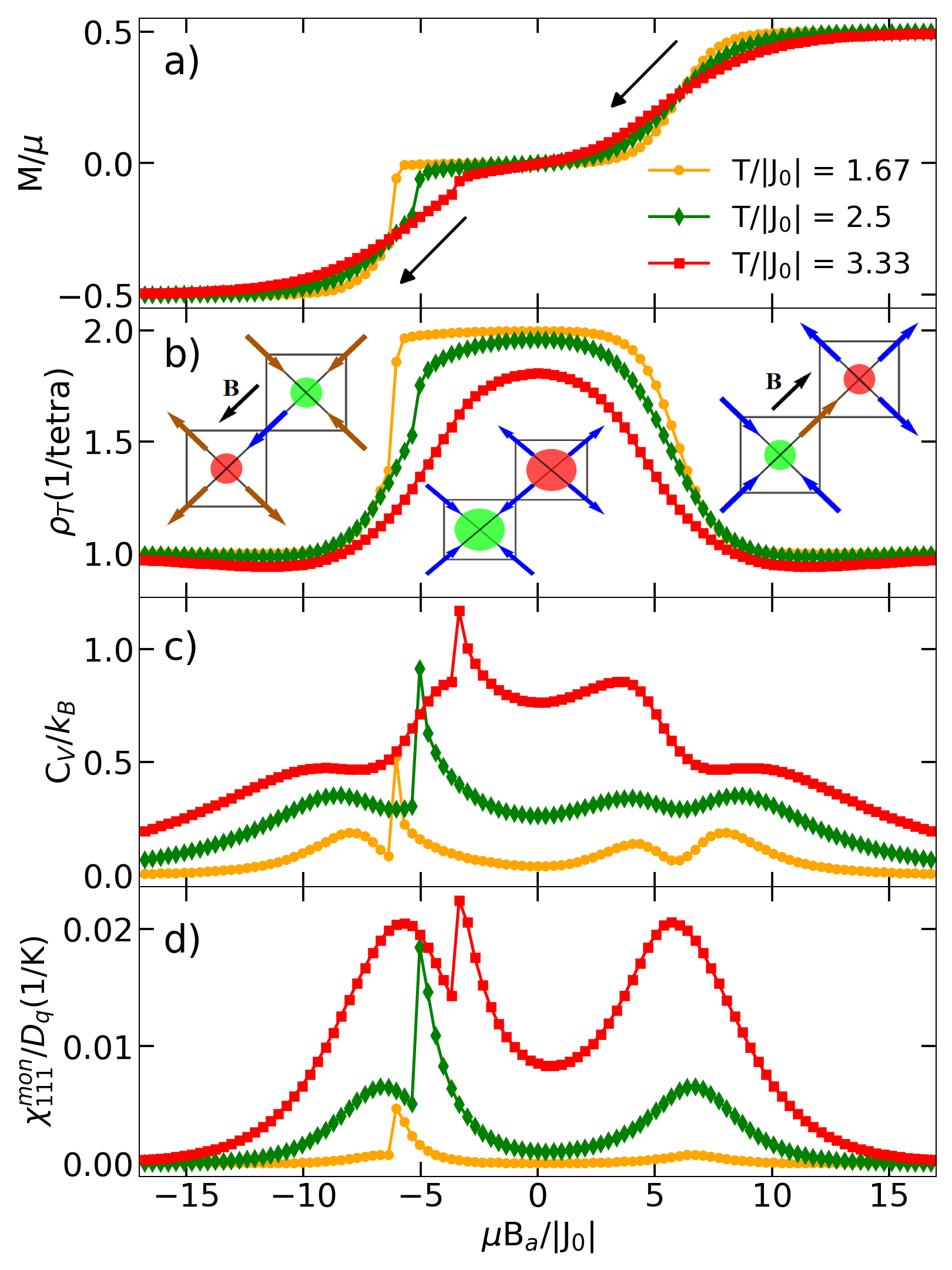}
    \caption{\textbf{$\bm{B} \parallel$ [111] magnetoelastic behaviour for the AIAO antiferromagnet.} Magnetization (a), density of monopoles (b), specific heat (c), and electric susceptibility due to monopoles (d) as a function of magnetic field for three different temperatures below $T_C$. The curves were measured for decreasing field (black arrows on Panel a), and show a clear asymmetry around $B=0$. The sharp features at $B<0$ correspond to phase transitions (see dashed orange lines in Fig.~\ref{fig:AIAO_B111_PhD}). Panel b) shows three schematic views of the spin configurations that are stable at low temperature in a two-dimensional projection. Spins can be divided into apical (colored in brown on the scheme on the right of Panel b), parallel to the applied field, and basal (colored blue on the same configuration).}
    \label{fig:AFB111} 
\end{figure}

The metamagnetic transition that takes place in Ising pyrochlores as a function of magnetic field $\bm{B} \parallel$[111] has been widely studied in the context of spin ices~\cite{fennell2002field,Sakakibara03,isakov2004magnetization,Castelnovo,molavian2009proposal,Borzi2016intermediate} and, more recently, for AIAO Ising antiferromagnets~\cite{lhotel2015fluctuations,tian2016field,opherden2017evolution,opherden2018inverted,xu2019anisotropic}. For this orientation, the field couples with all four spins in a tetrahedron. However, the so called ``apical'' spin (sitting on triangular planes) has its full component along ${\bm B}$, while the projection of the three ``basal'' spins along this direction is 1/3. 

Fig.~\ref{fig:AFB111} condenses the results of our Monte Carlo simulations for this field direction. The curves have been measured for decreasing magnetic field $B$, as indicated by the black arrows.  Panel a) shows the magnetisation curves for three different temperatures. For \textit{positive} fields we observe just a crossover on decreasing $B$; it links saturation at $M/\mu = 0.5$ (corresponding to the fully polarised crystal of single monopoles, schematized in the right inset to Panel b)) with the crystal of double monopoles (central inset to Panel b)), with $M=0$. Correspondingly, the monopole density (Panel b)) climbs smoothly from $1$ to $2$, and the specific heat (c)) shows two bumps defining a low valley at low temperature. Within this field range the difference between the energy cost for double and single monopoles is of the order of the thermal energy.

Naively, this smooth crossover is to be expected since \textit{i)} there is no spontaneous symmetry breaking and \textit{ii)} the formation of a crystal of double charges out of one of single ones involves no condensation energy in our nearest neighbour model.
It may then be surprising the behavior observed as the field is inverted. There is a sudden decrease in the magnetization and the density of monopoles near $\mu B/|J_0| \approx -5$, while the sharp peak in $C_V$ and the critical field where it occurs show finite scaling effects compatible with a first order transition  (not shown). 

The electric susceptibility due to monopoles at low $T$ provides us with some clues. We note that $\chi^{\rm mon}$ is very nearly zero at low temperature and positive fields (see curve at $T = 1~{\rm K}$ in Fig.~\ref{fig:AFB111}d)). We can understand this fact easily if we take into account that electric dipolar fluctuations would be mainly related with the fluctuations of apical spins (painted brown in the right inset to Panel b) in Fig.~\ref{fig:AFB111}). This is (again) a minority spin linking two single monopoles and thus involves a zero net electric dipole moment~\cite{khomskii2021electric}, and explains the almost zero electric susceptibility in spite of the obvious magnetic changes: the conversion of the single monopole crystal into an AIAO domain of a double monopole crystal involves only local, single spin flip events. On the other hand, as the field is reversed, this AIAO domain (with positive double charges in up tetrahedron) is then eventually transformed into a single monopole crystal with positive charges located in \textit{down} tetrahedra. As pointed out in Ref.~\onlinecite{xu2019anisotropic}, this requires the flipping of basal spins in each tetrahedron. It is a non-local event, since even if the flip of three basal spins may be an energetically favorable event in a given tetrahedron, there are other three adjacent tetrahedra where only a single spin has flipped.
Although the first spin flip links two tetrahedra with mutually canceling electric dipolar moments, other flips should lead to an uncompensated $P^{\OO}$ (i.e., to electric dipole fluctuations) and a measurable susceptibility $\chi_{\rm mon}$, explaining the sharp peak in Fig.~\ref{fig:AFB111}d) for $B<0$. The phase diagram for this phase transition, with its very particular hysteresis, will be discussed in more depth in Sec.~\ref{Sec:Discuss}.

\subsection{Comparison between different field directions}

\begin{figure}[htp]
    \includegraphics[width=0.80\columnwidth]{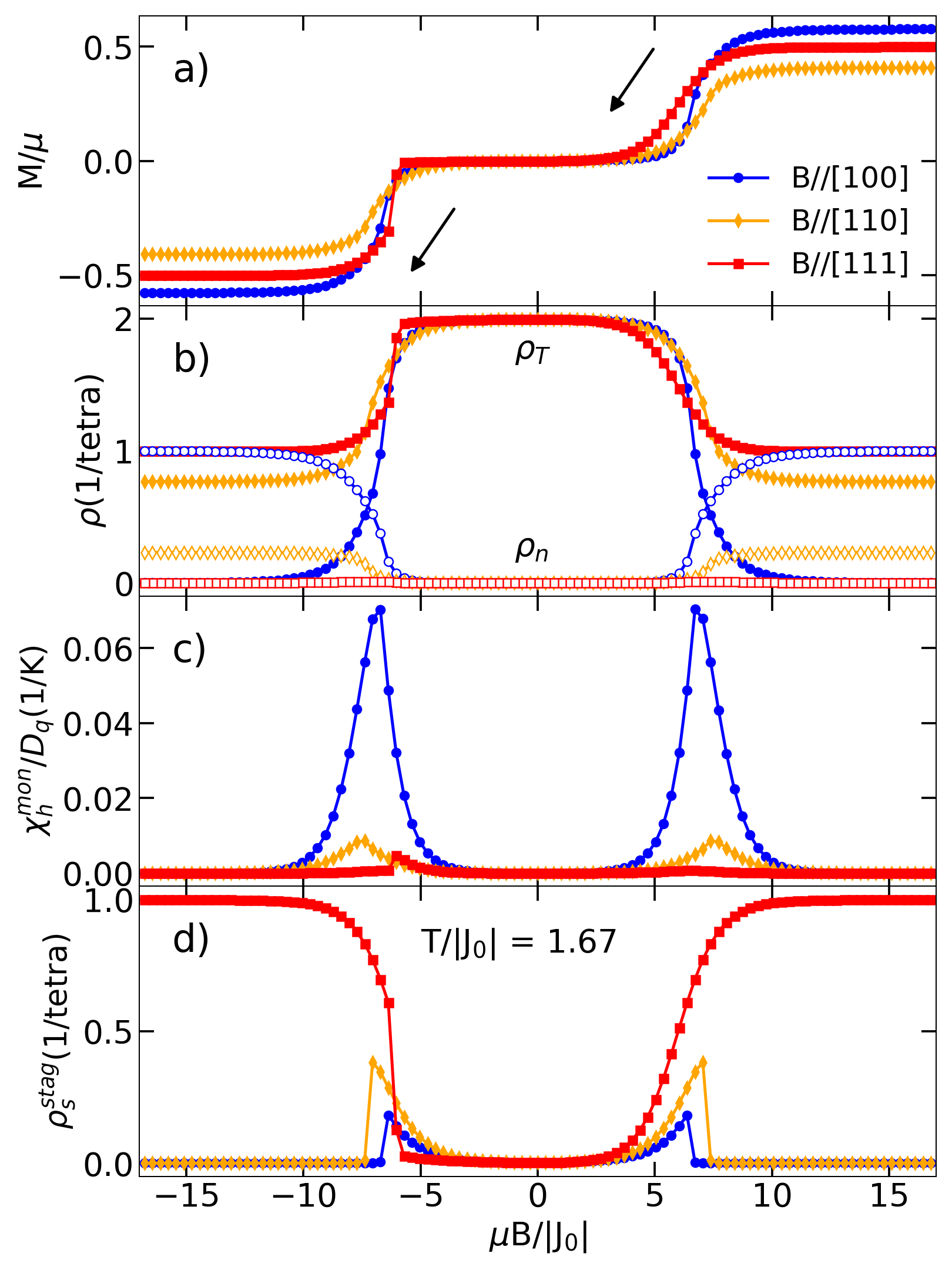}
    \caption{\textbf{Magnetoelastic behavior for the AIAO antiferromagnet with parallel electric and magnetic fields along different directions.} a) Magnetization , b) density of monopoles, c) electric susceptibility due to monopoles, and d) staggered charge density for double monopoles as a function of magnetic field for three different magnetic field directions at $T/|J_0|=1,67$. The curves were measured for decreasing field (black arrows on Panel a). Interestingly, $\chi^{\rm mon}_{\hat{h}}$ (Panel c)) is bigger for field orientations where the density of neutral sites (Panel b)) is bigger, and the staggered charge density (Panel d)) is smaller.}
    \label{fig:AFB} 
\end{figure}

As it happens with the ferromagnetic -- spin ice -- version, the antiferromagnetic AIAO phase also has an anisotropic response to a magnetic field.
Fig.~\ref{fig:AFB}a) shows the magnetization curves at $T/J_0=1.67~{\rm K}$ for three different magnetic field directions, measured again for decreasing fields. In red we re-plot the curve for [111] direction as a reference; among the three, it is the only one that is not symmetric under the inversion of $B$.

Fig.~\ref{fig:AFB}b) plots the total density of monopoles (full symbols, saturating in $2$ for a crystal of double charges), and that of neutral sites (hollow symbols). Since double monopoles have no magnetic moment, the different orientations reflect how these charges (stable at $B=0$) are replaced by other more energetically favorable ones under a magnetic field. As we have seen, the [111] direction favors single monopoles ($\rho_T \xrightarrow[|\mu B/J_0| \gg 1]{} 1$); on the contrary, $\bm{B}\parallel~\rm{[100]}$ stabilizes neutral (2in-2out) sites, depleting the lattice of all magnetic charges at high fields. The direction [110] is interesting: while the magnetic field treats single monopoles and neutral sites on equal footing, single monopoles are configurationally preferred by the exchange energy term in Eq.~\ref{eq:HMag}. This explains why at high [110] fields the system is dominated by single monopoles, but with a non-negligible fraction of neutral sites due to the relatively high temperature~\cite{guruciaga2019monte}. An additional thing to note is that, differently from [111], $\bm{B}\parallel$[110] does not impose global charge order. We discuss below the magnetoelectric effects using this information on the phases and magnetic charges evolution with magnetic field.

Fig.~\ref{fig:AFB}c) shows the magnetoelectric response $\chi^{mon}_{\hat{e}}$; in each curve the fluctuations of $P^\OO_{\hat{e}}$ where calculated using $\hat{e}\parallel \bm{B}$, mimicking experimental configurations used previously~\cite{Saito2005magnetodielectric}.
The  sharp peaks observed in $\chi^{mon}$ are related to phase transitions. The one for $\bm{B} \parallel$[111] at negative fields (red curve) has been already mentioned, and will be further discussed in Sec.~\ref{Sec:Discuss}. The symmetric transitions for [100] (blue) are in correspondence with the destruction of a Zn-blende charge crystal due to the proliferation of neutral sites. The orange curve corresponds to $\bm{B}\parallel$[110]. Here the double monopole crystal gives place to a single monopole liquid.  This happens after a narrow field range dominated by single charges of opposite sign held together by ``order by disorder''~\cite{Guruciaga16,guruciaga2019monte} ---see the staggered density of monopoles in panel d)--- . The relative peak height of the diverse field directions is perhaps perplexing: in spite of the fact that the only charges carrying an electric dipolar moment able to fluctuate are the single monopoles, the highest fluctuations occur for [100], where the lattice at each side of the transition point is mainly populated by double charges or neutral sites. 
This relies again on the fact that not only the density of single monopoles is important, but also
the electric dipolar correlations between neighboring single monopoles of opposite charge~\cite{khomskii2021electric}. The alternate charge order is enhanced by a [111] field, and by the presence of double monopoles. On the other hand, it is weakened by neutral sites~\cite{Guruciaga16}. Indeed, we can see in Fig.~\ref{fig:AFB} that the value of the staggered charge density for the different field directions (Panel d)) decreases with the density of neutral sites at the transition (Panel b)), while the electric response at the peak increases  with $\rho_{\rm n}$ (Panel c)). Although with a smaller $\rho_{\rm s}$, the presence of a high density of neutral sites favors a bigger peak in $\chi^{mon}$ for $\bm{B} \parallel$[100].

\section{Ferromagnetic, 2-in--2-out spin ice systems} \label{sec:Ferromag}

We now consider the ferromagnetic case, with $J_0 > 0$ in the magnetoelastic part of the MeSI Hamiltonian, Ec.~\ref{eq:HMag}. 
In this case, and in order to do a better comparison between our simulation and experimental results, we have also included long range dipolar interactions (Eq.~\ref{eq:dipm}) using the Ewald method~\cite{melko2004monte}, and exchange-like interactions from the second and the two types of third nearest neighbors:
\begin{equation}\label{eq:H_SpinIce}
    \begin{split}
    \mathcal{H} \ = \ \mathcal{H}^{\rm eff}_0 & + J_2 \sum_{\langle ij\rangle_2} S_i S_j + J_3 \sum_{\langle ij\rangle_3} S_i S_j + J_3^{'} \sum_{\langle ij\rangle_{3'}} S_i S_j \\
    & + \mathcal{H}^{{\tiny \rm Dip.M.}} + \mathcal{H}^{\rm Zeeman} \ .    
    \end{split}
\end{equation}
Here, $J_k$ is the $k$-th neighbor exchange constant, and $\langle \dots \rangle_k$ indicates summing over $k-$type neighbors. We use parameters for $\mu$, $D$ and $r_{nn}$ for \DTO\ taken from Ref.~\onlinecite{yavors2008dy}, and the optimized value of the exchange constants from Ref.~\onlinecite{Borzi2016intermediate}. We again assume $J_{ml}<J_0$; we will check the consistency of this assumption after comparing with the experimental results.

\subsection{Zero magnetic field}

\begin{figure}[htp]
    \includegraphics[width=0.8\columnwidth]{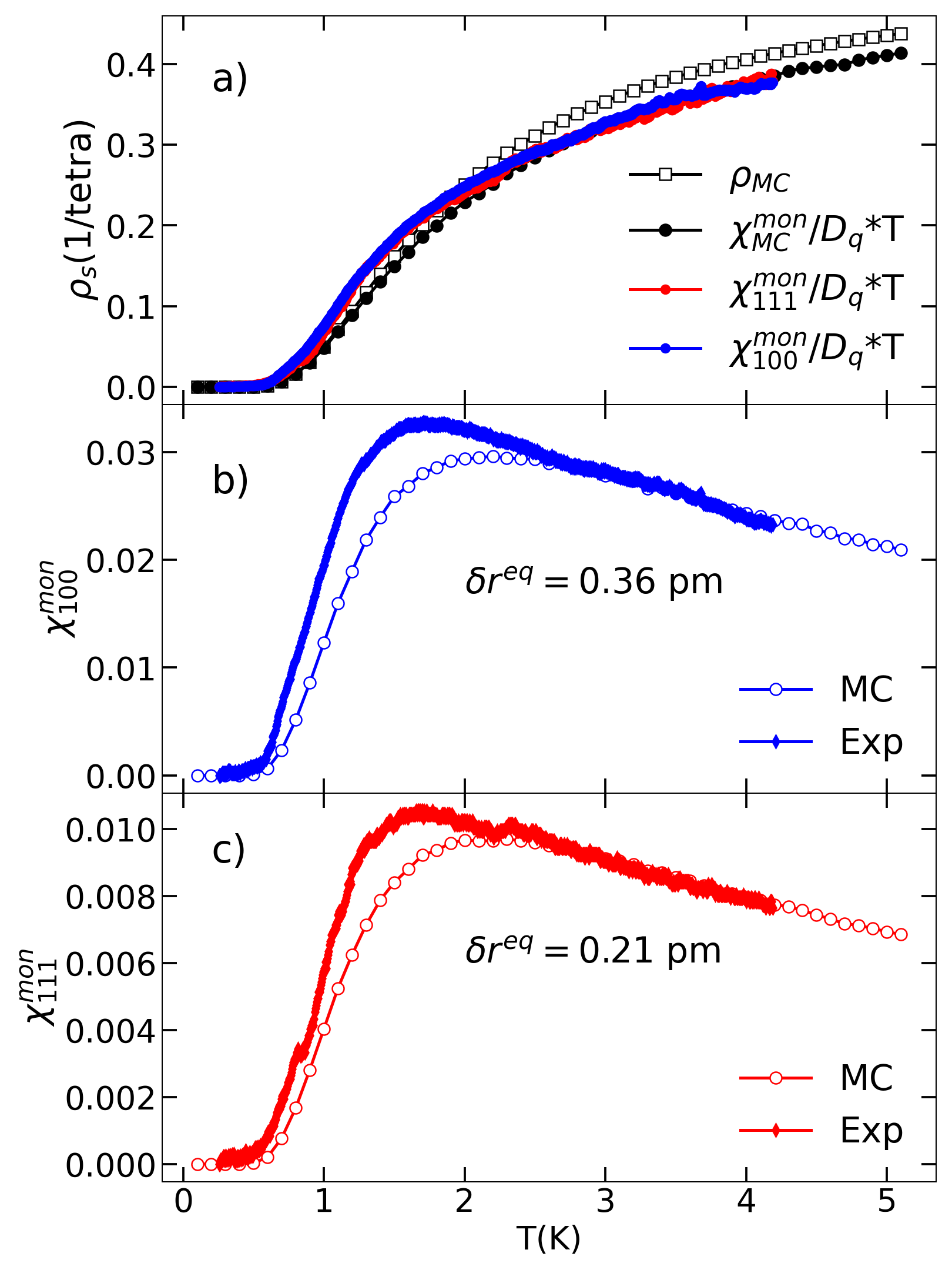}
    \caption{\textbf{Magnetodielectric behavior for \DTO~spin ice from Monte Carlo simulations (zero magnetic field), together with experimental measurements adapted from Saito et al\cite{Saito2005magnetodielectric}. } a) Single monopole density from simulations, compared with estimations using $\chi^{\rm mon}$ from MC simulations and experiments in the lower panels.  b) and c) Electric susceptibility due to monopoles, $\chi^{mon}$ for [100] (b) and [111] (c) electric field directions. Monte Carlo simulations (open circles) are compared with experimental data (full diamonds), after background subtraction. The shape of the experimental curves coincides for directions, as do the simulated ones, giving us confidence on the subtraction procedure. The value of the displacement $\delta r^{\rm eq}$ in the simulations for each direction, specified in the figures, was chosen so that the curves coincide in the high $T$ regime.}
    \label{fig:FerroFrusB0}
\end{figure}

The black curve with open symbols in Fig.~\ref{fig:FerroFrusB0}a) shows the density of single monopoles for \DTO\ as calculated from numerical simulations.  The number of monopoles is exponentially low at low temperatures, while the neutral background 
contributes to decorrelate them. In contrast with what we studied in Sec.~\ref{sec:AIAO} for the antiferromagnet (Fig.~\ref{fig:AFB=0}), we now expect correlation effects to increase with $T$. Figs.~\ref{fig:FerroFrusB0} b) and c) show the results of the simulated susceptibility (open circles) as a function of temperature for $B=0$ and for two different directions of the electric field $\hat{e}$ ; the method we used to calibrate the vertical axes will be made clear in the next paragraph, and will be further discussed in the Sec.~\ref{Sec:Discuss}. The maximum in the electric response due to monopoles is near $2~{\rm K}$ for both directions (in the absence of a symmetry breaking field, $\chi^{mon}_{\hat{e}}$ is isotropic), at higher temperatures than in the specific heat~\cite{Bramwell01,morris2009dirac}. As we did for the AIAO case, we can use this electric response at zero field to provide an estimate for the number of monopoles in the sample (black curve with full symbols in Fig.~\ref{fig:FerroFrusB0}). The estimate is indistinguishable from the simulated value at low temperature (low $\rho_{s}$), and remains within $10\%$ of the measured value within the inspected temperature range. We will now profit from the fact that spin ices are among the best studied frustrated materials and compare our simulations with previous experimental results in order to obtain quantitative information on the system.

Saito et al.~\cite{Saito2005magnetodielectric} measured the real and imaginary part of the magnetodielectric constant for \DTO\ by subtracting the contribution from the samples geometrical changes to the capacitance. Ignoring dynamical effects one would have naively expected the behavior of our simulated curves in Figs.~\ref{fig:FerroFrusB0} b) and c) to be similar to the real part of the electric susceptibility (Figs. 4 and 5 of Ref.~\onlinecite{Saito2005magnetodielectric}). On careful inspection, we see that although there are some common features, the similarity on the overall behavior is not so apparent.  The reason behind this contrast is of course that in a real sample there are other temperature-dependent contributions to the dielectric response aside from that coming from distortions associated to monopoles.  As we will argue, they can be in principle subtracted by using data measured at high magnetic fields.  

At high $B$ (such that $\bm{\mu}\cdot \bm{B} \gg T$ ) all magnetic moments should be saturated. No new electric dipolar moments related to monopoles are then created, and none fluctuate; all remaining contributions constitute then a background that may still depend on $T$, but which does not originate on magnetic monopoles, and can thus be subtracted from the curves at $B=0$. In our case, we used as background the curves from Ref.~\onlinecite{Saito2005magnetodielectric} measured as a function of temperature at the highest fields. In the subtraction we also added a constant to this background, to ensure that $\chi^{\rm mon}\to 0$ for $T\to 0$. This constant is very small (on the scale of the overall variation of $\chi^{\rm mon}$ with temperature) for $B\parallel$[100], and bigger than this scale (of the order of 0.024 in Fig. 5c)) for $B\parallel$[111] (see also Figs. 4 and Fig. 5 in Saito et al.~\cite{Saito2005magnetodielectric}).  The results are  displayed in Figs.~\ref{fig:FerroFrusB0} b) and c) on top of the numerical simulations (full circles).  
We can see that the resulting experimental curves are quite similar, peaking approximately at the same temperature. This is remarkable taking into account that this maximum was absent in the untreated data, and that both curves had different shapes. While there are some noticeable differences between the experiments and our simulations, the overall agreement is good, particularly considering the many approximations in the model, the subtraction method used with the experimental curves (that have already undergone a previous background subtraction), and the fact that these curves correspond to dynamic rather than static data.

For each of the two Monte Carlo curves in Figs.~\ref{fig:FerroFrusB0} b) and c) we determined a value for $D_q$ (i.e., a value for electric dipole moment $p_q$) so that they approximately coincide with the experimental ones from Saito et al. at high temperature. Assuming that the displaced charge is twice the electron charge~\onlinecite{Jaubert2015crystallog}, we obtain as estimations for the \OO\ displacement $\delta r^{eq} = 0.21~{\rm pm}$ for [111] and $\delta r^{eq} = 0.36~{\rm pm}$ for [100].  The difference in the values obtained for the different directions can be used as a way to estimate the magnitude of the error in its inferred value.  

We understand ours is the first direct measurement (in the sense that it is coming from an electric property) of the value of the dielectric dipole moment related to monopoles. These values are between $1/3$ and $1/2$ of that for the \OO\ displacement evaluated for \TTO, a compound which is known for its big magnetoelastic coupling~\cite{Fennell2014magnetoelastic,ruff2010magnetoelastics}. There are at least two previous, more indirect estimates for $p_q$ in \DTO. The first one~\cite{slobinsky2021monopole} corresponds to an \OO-ion displacement $\delta r^{eq} \approx 0.1~{\rm pm}$.  It is based on the independent estimate for $\tilde{\alpha}$ for \DTO\ from uniaxial pressure studies~\cite{edberg2019dipolar}, where they inferred the change in the exchange constants $J_0$ as a function of deformation; and also on the value of the elastic constants for \DTO\ from Raman, infrared spectroscopy and modelization~\cite{gupta2009lattice,kushwaha2017vibrational}. The second one, by Sarkar and Mukhopadhyay~\cite{sarkar2014dynamics}, is grounded on theoretical work on spin currents on noncolinear magnets~\cite{katsura2005spin}. Taking as inputs estimates for the hybridization energy between the Dy site and the O one involved in superexchange, they obtain $p_q \approx 10^{-30}~{\rm Cm}$. This value is approximately ten times what we deduce from magnetoelectrical experiments, and would imply an electric response linked to monopoles (proportional to $p_q^2$, Eq.~\ref{eq:chi_mon}) two orders of magnitude bigger than the experimental curves in Figs.~\ref{fig:FerroFrusB0}b) and c).

As we did for the antiferromagnet, we can now use the experimental and the theoretical curves to estimate the density of monopoles, assuming each tetrahedron contributes independently. This is shown in Fig.~\ref{fig:FerroFrusB0}a). The estimation is very good at low temperatures, with an error near 10\% in the high temperature limit.

There are polarization measurements under magnetic fields in \HTO~\cite{liu2013multiferroicity}, taken at temperatures above $2$~K. However, the data does not allow for a reliable comparison with our simulations, in particular since the temperature range explored excludes the maximum in the susceptibility.

\subsection{Magnetic field parallel to [111]}

\begin{figure}[htp]
    \includegraphics[width=0.8\columnwidth]{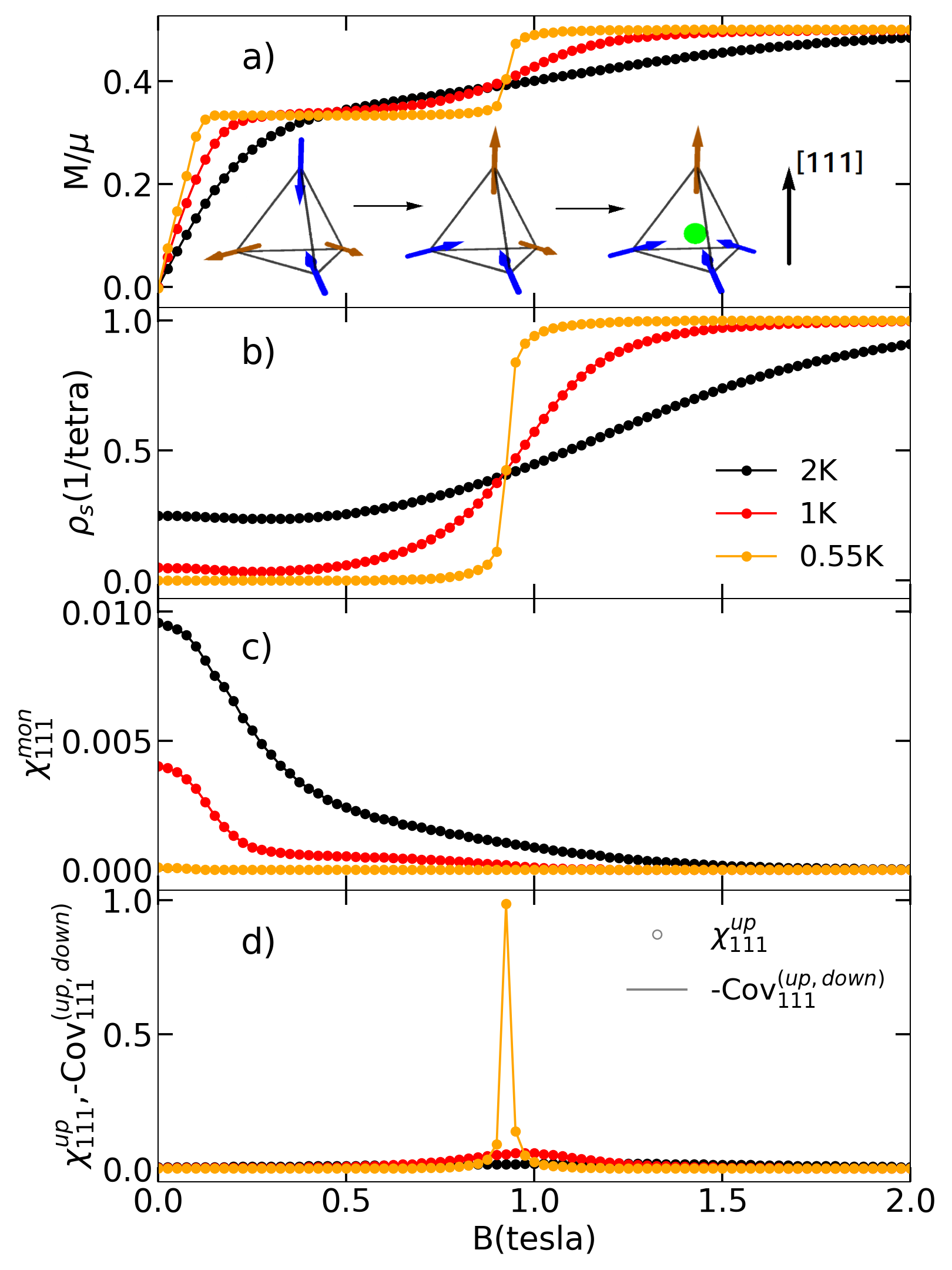}
    \caption{\textbf{Simulated magnetodielectric behaviour due to monopoles for \DTO~spin ice for $\bm{B} \parallel {\rm [111]}$.}  a) Magnetization vs. field at different temperatures above the critical point, with a sharp crossover at low $T$ near $0.9~{\rm tesla}$; b) The change into a crystal of single monopoles is shown at low temperatures by a sharp increase in the density of single monopoles, $\rho_{s}$. At low fields $\rho_s$ initially decreases with increasing field; this has a marked effect in the magnetodielectric response; c) Electric susceptibility due to monopoles as a function of field. It is nearly zero at low $T$, with a pronounced maximum near $B=0$ at higher $T$. There is no trace in $\chi^{\rm mon}$ of the sharp crossover near $0.9~{\rm tesla}$; d) Electric susceptibility due to monopoles on up tetrahedron only (full circles), and the negative of the covariance of the polarisation due to monopoles in up and down tetrahedra (lines). Both quantities are almost identical (and $\chi^{\rm down}_{111}=\chi^{\rm up}_{111}$) across the peak, explaining the absence of peak in $\chi^{\rm mon}$ at the metamagnetic transition. }
    \label{fig:SIB111}
\end{figure}

The presence of a magnetic field adds a layer of difficulty for a model to reproduce the experimental results. We will exemplify this for the case of $\bm{B}\parallel {\rm [111]}$. Fig.~\ref{fig:SIB111}a) shows the magnetisation for \DTO, simulated using the same model as in the previous subsection. Within the temperature regime $T \lesssim J_0 \approx 1{\rm K}$ and increasing $B$ there is a smooth crossover into the ``Kagome ice'' plateau, where the apical spins 
are fully polarized by the field. It is followed by a sharper evolution towards the saturated state near $0.9~{\rm tesla}$, marking the flipping of the basal spins (which are now the minority spins in each tetrahedron) in the Kagome planes. This crossover turns into a first order phase transition at $T$ below $\approx 0.4{\rm K}$ both for \DTO~\cite{Sakakibara03} and \HTO \cite{krey2012first}, a feature that is reproduced by the extended dipolar model we use~\cite{yavors2008dy,Borzi2016intermediate,samarakoon2022structural}. The curves for the density of single monopoles (panel b) have a decreasing behavior at low fields, followed by a sharper increasing one near $B=0.9~{\rm tesla}$: there, the symmetry breaking field destroys the monopole vacuum, and stabilizes a crystal of single monopoles. 

Our main subject of study here, the magnetodielectric response (Fig.~\ref{fig:SIB111}a), features a broad peak centered at $B=0$, quite noticeable at and above $1~{\rm K}$. It occurs in correspondence with the shallow maximum observed there in the density of single monopoles. Against our expectations, there is no trace of a peak nor any feature near $0.9~{\rm tesla}$ in the electric response $\chi^{\rm mon}_{\rm 111}$, in spite of the change in ground state from a vacuum to a crystal of single charges; this absent feature remains at even lower temperatures (not shown), where the change happens through a phase transition. 

Once again, the counter-intuitive fact of a reduction or total absence of an electric response when the density of single monopoles increases can be understood in terms of the monopole correlations we have discussed before. The progression from the Kagome plane towards saturation involves the flipping of minority spins (the basal ones, with fully polarized apical spins) that leads to pairs of single monopoles with opposing dipolar electric moments. Of course, other magnetic fluctuations aside from these do occur at finite $T$ (responsible for a finite $\chi^{\rm mon}$ near $0.9{\rm K}$ in Fig.~\ref{fig:SIB111}), but they do not generate any feature identifiable with the magnetic crystallization. This remarkable fact contrasts with the experimental measurements by Saito et al. for this field direction~\cite{Saito2005magnetodielectric}, evidencing a clear magnetodielectric peak near around $1~{\rm tesla}$ at all but the lowest temperatures ($0.26~{\rm K}$). Again we will stress that our model is only sensitive to dielectric changes in relation with magnetic monopoles, while Saito's samples should evidence those coming from any change concerning electric degrees of freedom within the crystal. It is thus reasonable to expect the full electric susceptibility to reflect the sharp changes taking place in a crystal with a changing symmetry breaking field applied and going across a sharp crossover/phase change ~\cite{stoter2019static}.

Regarding the broad maximum we observe in Fig.~\ref{fig:SIB111}c at low fields, this also seems to be absent from Saito's measurements (Fig. 6 in Ref.~\onlinecite{Saito2005magnetodielectric}); indeed, the real part of the dielectric constant as a function of $B$ at and below $0.55~{\rm K}$ has a depression at low fields. This is surprising in light of our previous success with the measurements at $B=0$ (Fig.~\ref{fig:FerroFrusB0}): it could be expected that the peak in the density of single monopoles centered at $B=0$ (with a very small degree of correlation) should make a contribution. A closer look to these curves reveal that at $T=1~{\rm K}$ the depression in the susceptibility at low fields has now leveled up with the rest of the curve, and there is even the hint of a peak at $T=2~{\rm K}$. We take this progression as the effect of the contribution of the single monopoles created on increasing temperature to the electric susceptibility at low fields. This seems to be qualitatively described by Fig.~\ref{fig:SIB111}c. Furthermore, such a peak at low fields (although much broader) is quite noticeable in the polarisation measurements performed in \HTO\ at $T=2~{\rm K}$. Finally, although the resolution in magnetic field is low and the temperature is relatively high, there does not seem to be any feature related to metamagnetism in these measurements (expected for this compound to be near $1.5~{\rm tesla}$). This coincides with the results of our simulations.

\section{Discussion}\label{Sec:Discuss}

\subsection{Magnetoelectric measurements, and the magnetic charge density}

Within the Results section, we have made use of magnetoelectric results in order to gain understanding on other quantities and, conversely, profited from other results to  enhance our understanding of magnetoelectric phenomena. The coupling between magnetic and electric degrees of freedom enables this exchange, and the MeSI model, in turn, provides a quantitative  connection between them.  This consideration assumes particular significance in scenarios where electric or magnetic measurements are easier to implement, or where the pursuit of an additional, parallel avenue to explore magnetic or electric phenomena proves advantageous.  Although not explored here, it is worth mentioning that this coupling could also prove a useful tool to control properties in a crossed way, allowing, for example, the manipulation of electric properties of a material by means of a magnetic field as in multiferroic materials. 

Since the proposal of magnetic monopoles in spin ice~\cite{Castelnovo}, there have been several proposals for indirect ways to measure the density of magnetic monopoles in spin-ice and other related materials. The methods involve measuring the specific heat, neutron scattering, magnetic noise, the response to an oscillating field, or (more recently) electronic magnetotransport (see Ref. ~\onlinecite{pearce2022monopoledensity} and references therein). Regarding this last suggestion, the electron scattering involves a magnetic channel (through the monopoles magnetic charge coupling with the electron spin) and an electric channel (through the monopoles electric dipolar moment and the electron charge). The studies in the previous sections indicate the possibility of using the electric dipolar moment carried by monopoles to measure the monopole density directly in electric polarization measurements~\cite{liu2013multiferroicity,lin2015experimental}, or through its fluctuations with magnetocapacitance experiments~\cite{Katsufuji2004magnetocapacitance,Saito2005magnetodielectric}.

Our results suggest that the method is as its best (see Figs.~\ref{fig:AFB=0} and~\ref{fig:FerroFrusB0}) for zero magnetic field and low temperatures ($T<J_0$) for spin ices. The contrast between simulations and experimental results seem to support the technique. Regarding AIAO antiferromagnets, the same method provides different information at two temperature regimes $B=0$. For $T$ above the ordering temperature the electric susceptibility due to monopoles, $\chi^{\rm mon}$, provides a reliable way to evaluate the density of single monopoles, $\rho_{\rm s}$. Below $T_C$, once the antiferromagnetic order is established, it conducts to the density of neutral sites, $\rho_{\rm n}$, connected with the density of single monopoles with uncorrelated electric moments.

\subsection{The phase diagram for the AIAO phase for $\textbf{B}\parallel$ [111]}

As mentioned before, there are previous reports~\cite{opherden2017evolution,opherden2018inverted,pearce2022monopoledensity,xu2019anisotropic} on the peculiar hysteresis for the AIAO antiferromagnet in a $\bm{B}\parallel$[111] magnetic field that we have measured here by means of the electric susceptibility $\chi^{mon}$ (Sec.~\ref{sec:AIAO}). However, to our knowledge there is no study of the full $B -T$ phase diagram.  We will undertake this task now using our magnetic model, which is perhaps the simplest possible in the pyrochlore lattice, with nearest neighbors interactions plus a Zeeman term. 

We start by identifying the magnetic states. With the magnetic field explicitly breaking the symmetry, we will talk of different phases if they are separated by singularities in the thermodynamic quantities. In this respect, finite size scaling reveals that the observed discontinuities in Fig.~\ref{fig:AFB111} are related to a true first order phase transition. Given the water-vapor transition found in spin ices for the same field direction~\cite{Sakakibara03}, it is tempting to associate the abrupt decrease of the total density of monopoles at low temperature (panel b in Fig.~\ref{fig:AFB111}) with a phase transition between the crystal of double monopoles and the Zn-blende crystal of single monopoles. However, this identification is challenged by the fact that the same figure evidences \textit{no} phase transition at positive fields. \textit{Only} if a domain type labeled by the sign of the average magnetic charge in up tetrahedra (let us say AIAO, with charge $+2Q$) at low fields is different from the one at higher field modulus (correspondingly, $-Q$ in up tetrahedra) we see evidence of a phase transition. This points to the first fact: there are only two different magnetic phases. As with the scalar magnetisation for an Ising ferromagnet, these phases are identifiable by the sign of the average charge in up tetrahedra, \textit{irrespective of its magnitude} (the crystal with $Q$ and $2Q$ in up tetrahedra correspond to the same phase, explaining the lack of a phase transition in Fig.~\ref{fig:AFB} for positive fields).

\begin{figure}[htp]
    \includegraphics[width=0.70\columnwidth]{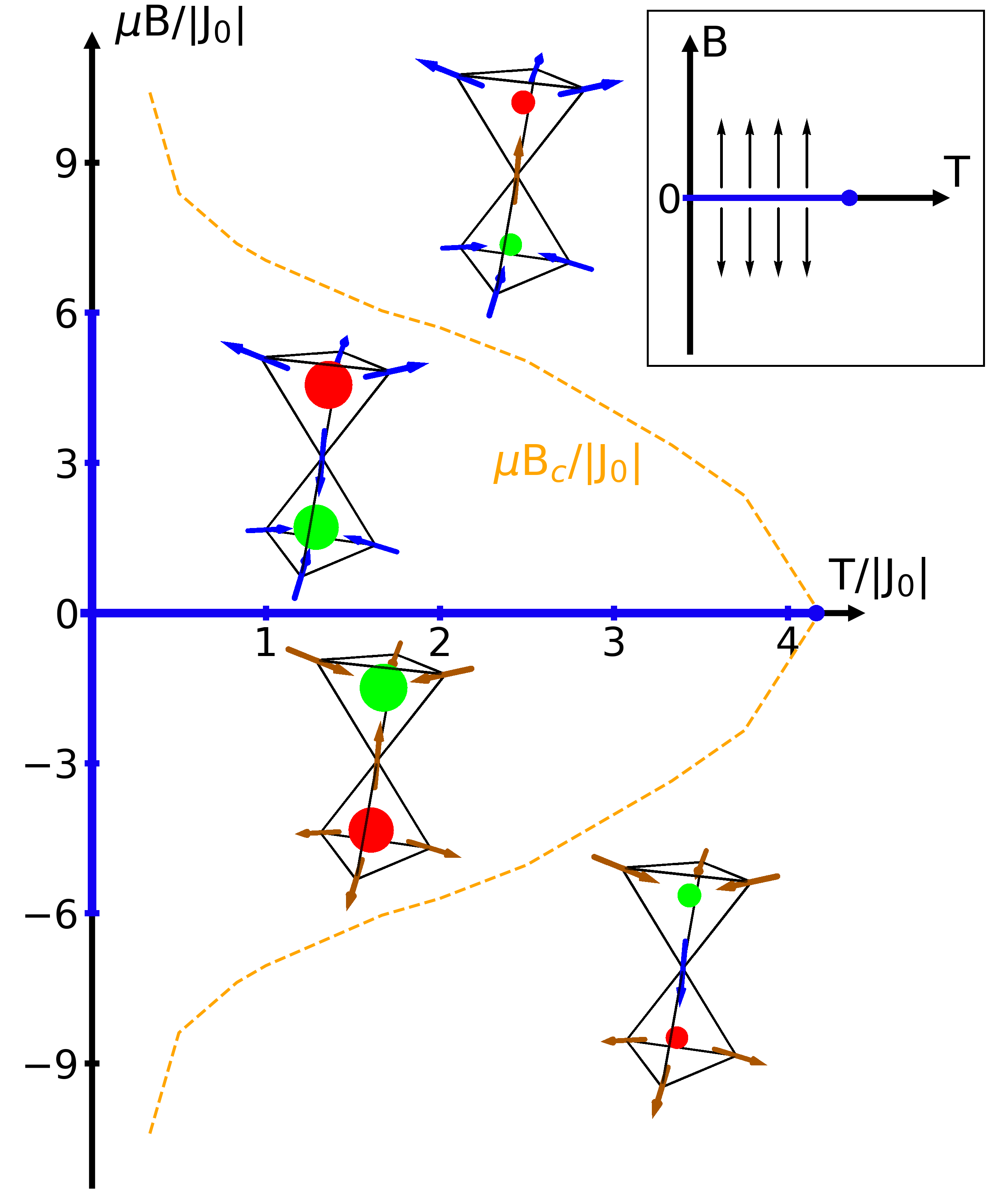}
    \caption{\textbf{Phase diagram for a simple AIAO antiferromagnet in a [111] field.}  The AIAO and AOAI phases at positive and negative fields are separated at non-zero temperatures by a coexistence line at $\mu B=0$ (blue horizontal line) terminating in a critical end-point indicated by a blue circle. The sign (and not the magnitude) of the average magnetic charge in each diamond sublattice is the feature that can be used to label each phase. Since a magnetic field selects a phase through their excitations, there is coexistence between these two phases at $T=0$ even at non-zero magnetic field. The features we observed for negative fields in Fig.~\ref{fig:AFB111} are related to the limit of phase metastability, indicated here by a dashed orange line. \textit{Inset:} Phase diagram for an Ising ferromagnet. The thick blue line for $B=0$ indicates the region of coexistence (a first order transition line) between the two polarized phases. Their distinction disappears above the Curie temperature, indicated by a blue circle. This phase diagram is similar --but not identical-- to that of the AIAO antiferromagnet.}
    \label{fig:AIAO_B111_PhD}
\end{figure}

The inset of Fig.~\ref{fig:AIAO_B111_PhD} shows the schematic phase diagram for a colinear Ising ferromagnet; a horizontal first order transition line ending on a terminal critical point separates the two phases, labeled by the sign of $M$. Only if the sample is polarized such that $M<0$ ($M>0$) at $B=0$ there will be a non analytic behavior at positive (negative) fields due to metastability. We now propose that a similar phase diagram is valid for the AIAO antiferromagnet (blue lines in Fig.~\ref{fig:AIAO_B111_PhD}, with stable configurations drawn schematically above and below the horizontal transition line). 

We should first address the question of the stability of the different domains in an applied field. Unlike the Ising ferromagnet, both the AIAO and AOAI domains (now \textit{phases}) have \textit{zero} magnetization. What can change their relative stability in an applied magnetic field? It is important to remember that, differently from \NIrO and \HoIrO ~\cite{tian2016field,pearce2022monopoledensity,ma2015mobile} there is no second magnetic lattice in this case. Furthermore, there is no spin canting in our model that may tilt the energetic balance towards one or the other phase~\cite{opherden2017evolution,opherden2018inverted}.
The answer to the previous question then rules out the differences between ground state configurations as the source of stability, and should then involve their respective excitations. 
Starting from the AIAO phase ($2Q>0$ in up tetrahedra), a field $B>0$ would favor leaving the three polarized basal spins as they are, and flipping the apical spin to make it fully parallel to the field ---and thus have two monopoles with maximum magnetic moment along [111] and a relatively big Zeeman energy reward ---see the central and top left tetrahedron in Fig.~\ref{fig:AIAO_B111_PhD})b). On the other hand, the AOAI phase in the same positive field should favor low energy excitations obtained after flipping a basal spin (with the apical fully polarized, but the other basals having a component against the field). This alternative excitations implies also creating two single monopoles, but has a lower Zeeman energy than the former, making the excited AOAI phase at a given $B>0$ less favored that the corresponding excited AIAO one.

This way to stability is reminiscent of the phenomenon of \textit{order by disorder}~\cite{chalker2011introduction}, in that the low energy excitations determine the smaller free energy among the available ground states.  A consequence of this is that the ground state is degenerate at $T=0$ even for non-zero fields, something we acknowledge with the vertical blue segment ending at $\pm 6J_0$ in Fig.~\ref{fig:AIAO_B111_PhD}b. 
The figure also shows an orange dashed curve, which marks the limit of metastability as measured in our simulations for a domain with positive (negative) charge in up tetrahedra with a negative (positive) field. It was constructed using the phase transition points in the $C_V$ vs. $B$ curves (Fig.~\ref{fig:AFB111}c) at each temperature and $L=4$. The curve has a a Gaussian shape; the critical end point peaks at $B=0$, near $T/|J_0|\approx 4.18$. On decreasing $T/|J_0|$ it first appears to intersect the axis $T=0$ at $\mu B/|J_0|=\pm 6$. However, at very low $T$ it flattens pushing the spinodal curve towards much higher fields. This is due to the lack of single monopole excitations at very low temperatures: without single monopoles the single spin flip dynamics raises the energy barrier to nucleate the stable domain to $\mu B/|J_0|=\pm 18$.

The physics studied in this subsection may have implicances in relation to previous studies.
As we said, this simple mechanism for stabilizing antiferromagnetic domains would be operative even in the absence of a second magnetic lattice in the pyrochlore lattice, and without spin canting. It may thus be at work in experiments of antiferromagnetic domain handling where other mechanisms involving other degrees of freedom or energy terms have been contemplated~\cite{opherden2017evolution,opherden2018inverted,tian2016field,pearce2022monopoledensity,ma2015mobile}. Using the language of Ref.~\onlinecite{pearce2022monopoledensity} we can summarize the effect saying that the single monopolar excitations of the AIAO or AOAI phases exert a pressure on \textit{its own} antiferromagnetic background.

\subsection{Estimation of the electric dipole moment $p_q$, distortion $\delta r_{\rm eq}$, and $J_{ml}$ in \DTO}

In spite of its simplicity and the number of approximations, the quantitative comparison of our simple MeSI model with experimental results for spin ices physics brought a number of results worth discussing. The first one regards the value of the dipolar electric moment for \DTO, $p_q = (1 \pm 0.2) \times 10^{-31}{\rm Cm}$ and the central \OO-ion displacement $\delta r^{\rm eq} = 0.28 \pm 0.07$~pm. Since the prediction of an electric dipole moment accompanying single magnetic monopoles, there have been a number of experimental results connecting different effects with the distortion $\delta r_{\rm eq}$~\cite{grams2014critical,jin2020experimental,Jaubert2015crystallog,slobinsky2021monopole}. To our knowledge, ours is the first direct estimation of its value.

We stress now that the real part of the dielectric constant as a function of temperature measured at zero field in Ref.~\onlinecite{Saito2005magnetodielectric} has different shapes for $\hat{e}\parallel$[100] and [111].  The same is true for the two high field backgrounds at $B \approx 6$~tesla. In spite of this, it is quite remarkable that after subtraction the shape of the experimental electric susceptibility look essentially the same (full circles in Figs.~\ref{fig:FerroFrusB0} b) and c)). This makes us confident on the subtraction procedure. On the other hand, and in spite of this coincidence, there is an overall scale factor between both experimental curves which is very near $3$. This unexpected factor is the main source of error in the estimation of the magnetoelectric parameters for \DTO.

One of the assumptions along our work was that the effective magnetoelastic energy in the form of the four spin term in Eq.~\ref{eq:HMag} could be neglected compared with the other contributions. We can show the consistency of this for \DTO\ in different, independent ways. There are previous studies of this material where the evolution of the exchange constants upon uniaxial pressure was measured~\cite{edberg2019dipolar}. This, together with the geometrical changes taking place in the unit cell allow us to estimate $\Tilde{\alpha} \approx 40~{\rm K}$ for \DTO. Using expression~\ref{eq:deltaRmin} together with the average value we obtained from Figs.~\ref{fig:FerroFrusB0}b and c of $\delta r^{\rm eq} \approx 0.3~{\rm pm}$, we obtain $J_{ml} \approx 0.03~{\rm K} \ll J_0$. Otherwise, we can use the interatomic force constants inferred from infrared measurements as a direct estimation for the elastic constant, obtaining $K \approx 3\times 10^5~{\rm K}$. With this value and our estimate for $\delta r^{\rm eq}$ we obtain $J_{ml} \approx 0.05~{\rm K}$. Another consistency check regards the importance of the electric dipolar interactions between monopoles. Using the two values obtained for $\delta r^{\rm eq}$ we estimate electric dipolar energies for neighboring monopoles ranging between $0.03$ and $0.1$~K. Although small, the second estimation is comparable to the third nearest neighbors magnetic dipolar interactions in \DTO, about $0.18$~K.

In order to keep things simpler we have assumed a crystal with no imperfections. However, a static distortion caused by defects (for instance, O deficiency~\cite{prabhakaran2011crystal}) can affect what we want to simulate if it alters the probability of occurrence of spin configurations with different associated electric dipoles (for example, if it modifies the exchange constants). Impurities replacing ions can change the magnetic energy balance and, more drastically, the local symmetry, leading to electric dipoles. Since the experiments we compare with here are performed on single crystals, we expect defects associated to grain boundaries not to be dominant in this case.

\subsection{Correlations and in-field measurements for spin ice}

At non-zero field, the comparison between our results for the spin ice \DTO\ and the experiments seem to show a considerably poorer agreement than for $B=0$. The existence of a varying magnetisation $M(T,B_e)$ due to an applied magnetic field seems to have a big influence on the dielectric properties, beyond the direct contribution related to magnetic monopoles. This is made apparent by the fact that the MeSI model predicts no peak in $\chi^{mon}$ for the metamagnetic transition in spin ices for $\bm{B}\parallel$[111] (Fig.~\ref{fig:SIB111}), while measurements of the ac-dielectric constant evidence a big peak near the transition for \DTO~\cite{Saito2005magnetodielectric}. While it is likely that this lack of correspondence will be also observed in other compounds, there are measurements of the magnetocapacitance at $1.8~{\rm K}$~\cite{Katsufuji2004magnetocapacitance} and electric polarisation at $2~{\rm K}$~\cite{liu2013multiferroicity} in \HTO\ that seem to show no peak around $B=1.5~{\rm tesla}$. 

We will now stress an important aspect in association with these results. If we separate the contributions to the electric dipolar moment from up and down diamond sublattices (see Eq.~\ref{eq:p_q}) we can write the electrical susceptibility due to monopoles as,
\begin{align}
    {\chi^{\rm mon}_{\hat{e}}} \ = \ \chi^{\rm up}_{\hat{e}} + \chi^{\rm down}_{\hat{e}}+ 2\, {\rm Cov}^{({\rm up,down})}_{\hat{e}} \ ,
\end{align}
with
\begin{align}
    {\rm Cov}^{({\rm up,down})}_{\hat{e}} \ \equiv \ D_q \frac{3}{T N_q} \ \big( \ \big\langle {P}^{\rm up}_{\hat{e}} {P}^{\rm down}_{\hat{e}}  \big\rangle - \big\langle {P}^{\rm up}_{\hat{e}}  \big\rangle \big\langle {P}^{\rm down}_{\hat{e}}  \big\rangle \ \big) \ .
    \label{ec:covar}
\end{align}
It is to be noted that the electric susceptibility associated to fluctuations of dipoles in any of the sublattices is huge along the crossover from the Kagomé ice to the crystal of single monopoles (see the relative scales for the vertical axis in panels c) and d) in Fig.~\ref{fig:SIB111}). The very small susceptibility at $\approx 0.9$ tesla in Panel c) is what remains of the subtraction between two big terms ---the second of which is the covariance from Eq.~\ref{ec:covar}; it is surprising 
that any feature related with the crossover is removed from the electric response.
However, any influence on this correlation that is not taken into account by our simple model would imply a sizable contribution to $\chi^{\rm mon}$. Perhaps such an effect is behind the peak observed in the experiments by Saito et al.~\cite{Saito2005magnetodielectric} in association with metamagnetism for $\bm{B}\parallel$[111].


\section{Summary and Conclusions}\label{sec:Summ}

In this paper we have studied the magnetoelectric response of Ising pyrochlores, both the ordered antiferromagnetic ``all-in/all-out'' and the frustrated ferromagnetic ``spin-ice'' case. We worked in the limit of weak coupling between the magnetic and elastic degrees of freedom. For this, we have used numerical simulations based on the Magnetoelastic Spin Icse (MeSI) model, which considers the interplay between oxygen distortions and superexchange magnetic interactions in pyrochlores. We showed that the model can be simplified into a purely magnetic effective model. In this way, one can overcome the need to simultaneously simulate magnetic and elastic degrees of freedom, and thus gain a significant numerical benefit. Furthermore, the treatment supplies a unified framework that encompasses previous works~\cite{Jaubert2015crystallog}.
This streamlining opens up the potential for quantitative comparisons between simulations and experiments, allowing for a direct estimation of quantities such as the monopoles electric dipole moment $p_q$, and the local distortion $\delta^{\rm eq}$ associated to it for the spin-ice materials. The values obtained for \DTO\ using the experimental data from Saito et al.~\cite{Saito2005magnetodielectric} are $p_q = (1 \pm 0.2) \times 10^{-31}{\rm Cm}$ and the central \OO-ion displacement $\delta r^{\rm eq} = 0.28 \pm 0.07$~pm. This allows in turn for an estimation of new magnetoelastic and electric dipolar contributions that may affect in unsuspected ways the physics of this material. We found that both the four-spin term in Eq.~\ref{eq:HMag} and the dipolar interactions are small, but that the contributions from the latter are of the same order of magnitude as the magnetic dipolar interactions between third nearest-neighbors (and thus, not negligible).

Due to the electric dipole attached to monopoles, electrical properties can be used to investigate magnetic properties, or the establishment of magnetic order. As an example of the former, we have shown that magnetoelectricity provides a direct method to estimate the density of monopolar excitations in Ising pyrochlores. Furthermore, the magnetoelectric response showed a high sensitivity to monopole correlations; the reduced susceptibility due to this factor~\onlinecite{khomskii2021electric} lead many times to somewhat counterintuitive results. Regarding the latter possibility, we discussed the phase diagram of the antiferromagnetic pyrochlore under applied magnetic field along the [111] direction, and found a somewhat unusual situation where order at non-zero temperatures is stabilized not by the in-field energy related to the different ground state configurations, but by their relative accessibility to low energy fluctuations.   

At a broader level, our study highlights the possibilities opened up by the interplay between electrical and magnetic degrees of freedom, not only in terms of allowing for diverse probes into physical phenomena but also providing avenues to cross control electric and magnetic phenomena.


\begin{acknowledgments}
We would like to acknowledge useful discussions with T. S. Grigera. This work was supported by Consejo Nacional de Investigaciones Cient\'\i ficas y T\'ecnicas (CONICET) and Agencia Nacional de Promoci\'on Cient\'\i fica y Tecnol\'ogica (ANPCyT) Argentina, through PICT 2017-2347.

\end{acknowledgments}


\bibliography{biblio}

\begin{thebibliography}{58}%
\makeatletter
\providecommand \@ifxundefined [1]{%
 \@ifx{#1\undefined}
}%
\providecommand \@ifnum [1]{%
 \ifnum #1\expandafter \@firstoftwo
 \else \expandafter \@secondoftwo
 \fi
}%
\providecommand \@ifx [1]{%
 \ifx #1\expandafter \@firstoftwo
 \else \expandafter \@secondoftwo
 \fi
}%
\providecommand \natexlab [1]{#1}%
\providecommand \enquote  [1]{``#1''}%
\providecommand \bibnamefont  [1]{#1}%
\providecommand \bibfnamefont [1]{#1}%
\providecommand \citenamefont [1]{#1}%
\providecommand \href@noop [0]{\@secondoftwo}%
\providecommand \href [0]{\begingroup \@sanitize@url \@href}%
\providecommand \@href[1]{\@@startlink{#1}\@@href}%
\providecommand \@@href[1]{\endgroup#1\@@endlink}%
\providecommand \@sanitize@url [0]{\catcode `\\12\catcode `\$12\catcode
  `\&12\catcode `\#12\catcode `\^12\catcode `\_12\catcode `\%12\relax}%
\providecommand \@@startlink[1]{}%
\providecommand \@@endlink[0]{}%
\providecommand \url  [0]{\begingroup\@sanitize@url \@url }%
\providecommand \@url [1]{\endgroup\@href {#1}{\urlprefix }}%
\providecommand \urlprefix  [0]{URL }%
\providecommand \Eprint [0]{\href }%
\providecommand \doibase [0]{http://dx.doi.org/}%
\providecommand \selectlanguage [0]{\@gobble}%
\providecommand \bibinfo  [0]{\@secondoftwo}%
\providecommand \bibfield  [0]{\@secondoftwo}%
\providecommand \translation [1]{[#1]}%
\providecommand \BibitemOpen [0]{}%
\providecommand \bibitemStop [0]{}%
\providecommand \bibitemNoStop [0]{.\EOS\space}%
\providecommand \EOS [0]{\spacefactor3000\relax}%
\providecommand \BibitemShut  [1]{\csname bibitem#1\endcsname}%
\let\auto@bib@innerbib\@empty
\bibitem [{\citenamefont {Diep}(2013)}]{Diep2013frustrated}%
  \BibitemOpen
  \bibfield  {author} {\bibinfo {author} {\bibfnamefont {H.~T.}\ \bibnamefont
  {Diep}},\ }\href@noop {} {\emph {\bibinfo {title} {Frustrated spin
  systems}}}\ (\bibinfo  {publisher} {World Scientific},\ \bibinfo {year}
  {2013})\BibitemShut {NoStop}%
\bibitem [{\citenamefont {Lacroix}\ \emph {et~al.}(2011)\citenamefont
  {Lacroix}, \citenamefont {Mendels},\ and\ \citenamefont
  {Mila}}]{Lacroix2011introduction}%
  \BibitemOpen
  \bibfield  {author} {\bibinfo {author} {\bibfnamefont {C.}~\bibnamefont
  {Lacroix}}, \bibinfo {author} {\bibfnamefont {P.}~\bibnamefont {Mendels}}, \
  and\ \bibinfo {author} {\bibfnamefont {F.}~\bibnamefont {Mila}},\ }\href@noop
  {} {\emph {\bibinfo {title} {Introduction to frustrated magnetism: materials,
  experiments, theory}}},\ Vol.\ \bibinfo {volume} {164}\ (\bibinfo
  {publisher} {Springer Science \& Business Media},\ \bibinfo {year}
  {2011})\BibitemShut {NoStop}%
\bibitem [{\citenamefont {Ramirez}(1994)}]{Ramirez1994review}%
  \BibitemOpen
  \bibfield  {author} {\bibinfo {author} {\bibfnamefont {A.~P.}\ \bibnamefont
  {Ramirez}},\ }\href
  {https://www.annualreviews.org/doi/10.1146/annurev.ms.24.080194.002321}
  {\bibfield  {journal} {\bibinfo  {journal} {Annu. Rev. Mater. Sci.}\ }\textbf
  {\bibinfo {volume} {24}},\ \bibinfo {pages} {453} (\bibinfo {year}
  {1994})}\BibitemShut {NoStop}%
\bibitem [{\citenamefont {Moessner}\ and\ \citenamefont
  {Ramirez}(2006)}]{Moessner2006review}%
  \BibitemOpen
  \bibfield  {author} {\bibinfo {author} {\bibfnamefont {R.}~\bibnamefont
  {Moessner}}\ and\ \bibinfo {author} {\bibfnamefont {A.~P.}\ \bibnamefont
  {Ramirez}},\ }\href
  {https://physicstoday.scitation.org/doi/10.1063/1.2186278} {\bibfield
  {journal} {\bibinfo  {journal} {Physics Today}\ }\textbf {\bibinfo {volume}
  {59}},\ \bibinfo {pages} {24} (\bibinfo {year} {2006})}\BibitemShut {NoStop}%
\bibitem [{\citenamefont {Bramwell}\ and\ \citenamefont
  {Gingras}(2001)}]{Bramwell01}%
  \BibitemOpen
  \bibfield  {author} {\bibinfo {author} {\bibfnamefont {S.~T.}\ \bibnamefont
  {Bramwell}}\ and\ \bibinfo {author} {\bibfnamefont {M.~J.~P.}\ \bibnamefont
  {Gingras}},\ }\href {\doibase 10.1126/science.1064761} {\bibfield  {journal}
  {\bibinfo  {journal} {Science}\ }\textbf {\bibinfo {volume} {294}},\ \bibinfo
  {pages} {1495} (\bibinfo {year} {2001})}\BibitemShut {NoStop}%
\bibitem [{\citenamefont {Udagawa}\ \emph {et~al.}(2021)\citenamefont
  {Udagawa}, \citenamefont {Jaubert} \emph {et~al.}}]{udagawa2021spin}%
  \BibitemOpen
  \bibfield  {author} {\bibinfo {author} {\bibfnamefont {M.}~\bibnamefont
  {Udagawa}}, \bibinfo {author} {\bibfnamefont {L.}~\bibnamefont {Jaubert}},
  \emph {et~al.},\ }\href@noop {} {\emph {\bibinfo {title} {Spin Ice}}}\
  (\bibinfo  {publisher} {Springer},\ \bibinfo {year} {2021})\BibitemShut
  {NoStop}%
\bibitem [{\citenamefont {Prabhakaran}\ and\ \citenamefont
  {Boothroyd}(2011)}]{prabhakaran2011crystal}%
  \BibitemOpen
  \bibfield  {author} {\bibinfo {author} {\bibfnamefont {D.}~\bibnamefont
  {Prabhakaran}}\ and\ \bibinfo {author} {\bibfnamefont {A.}~\bibnamefont
  {Boothroyd}},\ }\href@noop {} {\bibfield  {journal} {\bibinfo  {journal}
  {Journal of Crystal Growth}\ }\textbf {\bibinfo {volume} {318}},\ \bibinfo
  {pages} {1053} (\bibinfo {year} {2011})}\BibitemShut {NoStop}%
\bibitem [{\citenamefont {Melko}\ and\ \citenamefont
  {Gingras}(2004)}]{melko2004monte}%
  \BibitemOpen
  \bibfield  {author} {\bibinfo {author} {\bibfnamefont {R.~G.}\ \bibnamefont
  {Melko}}\ and\ \bibinfo {author} {\bibfnamefont {M.~J.}\ \bibnamefont
  {Gingras}},\ }\href@noop {} {\bibfield  {journal} {\bibinfo  {journal}
  {Journal of Physics: Condensed Matter}\ }\textbf {\bibinfo {volume} {16}},\
  \bibinfo {pages} {R1277} (\bibinfo {year} {2004})}\BibitemShut {NoStop}%
\bibitem [{\citenamefont {Yavors’kii}\ \emph {et~al.}(2008)\citenamefont
  {Yavors’kii}, \citenamefont {Fennell}, \citenamefont {Gingras},\ and\
  \citenamefont {Bramwell}}]{yavors2008dy}%
  \BibitemOpen
  \bibfield  {author} {\bibinfo {author} {\bibfnamefont {T.}~\bibnamefont
  {Yavors’kii}}, \bibinfo {author} {\bibfnamefont {T.}~\bibnamefont
  {Fennell}}, \bibinfo {author} {\bibfnamefont {M.~J.~P.}\ \bibnamefont
  {Gingras}}, \ and\ \bibinfo {author} {\bibfnamefont {S.~T.}\ \bibnamefont
  {Bramwell}},\ }\href@noop {} {\bibfield  {journal} {\bibinfo  {journal}
  {Physical review letters}\ }\textbf {\bibinfo {volume} {101}},\ \bibinfo
  {pages} {037204} (\bibinfo {year} {2008})}\BibitemShut {NoStop}%
\bibitem [{\citenamefont {Borzi}\ \emph {et~al.}(2016)\citenamefont {Borzi},
  \citenamefont {G{\'o}mez~Albarrac{\'\i}n}, \citenamefont {Rosales},
  \citenamefont {Rossini}, \citenamefont {Steppke}, \citenamefont
  {Prabhakaran}, \citenamefont {Mackenzie}, \citenamefont {Cabra},\ and\
  \citenamefont {Grigera}}]{Borzi2016intermediate}%
  \BibitemOpen
  \bibfield  {author} {\bibinfo {author} {\bibfnamefont {R.~A.}\ \bibnamefont
  {Borzi}}, \bibinfo {author} {\bibfnamefont {F.~A.}\ \bibnamefont
  {G{\'o}mez~Albarrac{\'\i}n}}, \bibinfo {author} {\bibfnamefont {H.~D.}\
  \bibnamefont {Rosales}}, \bibinfo {author} {\bibfnamefont {G.~L.}\
  \bibnamefont {Rossini}}, \bibinfo {author} {\bibfnamefont {A.}~\bibnamefont
  {Steppke}}, \bibinfo {author} {\bibfnamefont {D.}~\bibnamefont
  {Prabhakaran}}, \bibinfo {author} {\bibfnamefont {A.~P.}\ \bibnamefont
  {Mackenzie}}, \bibinfo {author} {\bibfnamefont {D.~C.}\ \bibnamefont
  {Cabra}}, \ and\ \bibinfo {author} {\bibfnamefont {S.~A.}\ \bibnamefont
  {Grigera}},\ }\href {https://www.nature.com/articles/ncomms12592} {\bibfield
  {journal} {\bibinfo  {journal} {Nat. Commun.}\ }\textbf {\bibinfo {volume}
  {7}},\ \bibinfo {pages} {12592} (\bibinfo {year} {2016})}\BibitemShut
  {NoStop}%
\bibitem [{\citenamefont {Henelius}\ \emph {et~al.}(2016)\citenamefont
  {Henelius}, \citenamefont {Lin}, \citenamefont {Enjalran}, \citenamefont
  {Hao}, \citenamefont {Rau}, \citenamefont {Altosaar}, \citenamefont
  {Flicker}, \citenamefont {Yavors'kii},\ and\ \citenamefont
  {Gingras}}]{henelius2016refrustration}%
  \BibitemOpen
  \bibfield  {author} {\bibinfo {author} {\bibfnamefont {P.}~\bibnamefont
  {Henelius}}, \bibinfo {author} {\bibfnamefont {T.}~\bibnamefont {Lin}},
  \bibinfo {author} {\bibfnamefont {M.}~\bibnamefont {Enjalran}}, \bibinfo
  {author} {\bibfnamefont {Z.}~\bibnamefont {Hao}}, \bibinfo {author}
  {\bibfnamefont {J.~G.}\ \bibnamefont {Rau}}, \bibinfo {author} {\bibfnamefont
  {J.}~\bibnamefont {Altosaar}}, \bibinfo {author} {\bibfnamefont
  {F.}~\bibnamefont {Flicker}}, \bibinfo {author} {\bibfnamefont
  {T.}~\bibnamefont {Yavors'kii}}, \ and\ \bibinfo {author} {\bibfnamefont
  {M.~J.~P.}\ \bibnamefont {Gingras}},\ }\href@noop {} {\bibfield  {journal}
  {\bibinfo  {journal} {Physical Review B}\ }\textbf {\bibinfo {volume} {93}},\
  \bibinfo {pages} {024402} (\bibinfo {year} {2016})}\BibitemShut {NoStop}%
\bibitem [{\citenamefont {Samarakoon}\ \emph {et~al.}(2020)\citenamefont
  {Samarakoon}, \citenamefont {Barros}, \citenamefont {Li}, \citenamefont
  {Eisenbach}, \citenamefont {Zhang}, \citenamefont {Ye}, \citenamefont
  {Sharma}, \citenamefont {Dun}, \citenamefont {Zhou}, \citenamefont {Grigera}
  \emph {et~al.}}]{samarakoon2020machine}%
  \BibitemOpen
  \bibfield  {author} {\bibinfo {author} {\bibfnamefont {A.~M.}\ \bibnamefont
  {Samarakoon}}, \bibinfo {author} {\bibfnamefont {K.}~\bibnamefont {Barros}},
  \bibinfo {author} {\bibfnamefont {Y.~W.}\ \bibnamefont {Li}}, \bibinfo
  {author} {\bibfnamefont {M.}~\bibnamefont {Eisenbach}}, \bibinfo {author}
  {\bibfnamefont {Q.}~\bibnamefont {Zhang}}, \bibinfo {author} {\bibfnamefont
  {F.}~\bibnamefont {Ye}}, \bibinfo {author} {\bibfnamefont {V.}~\bibnamefont
  {Sharma}}, \bibinfo {author} {\bibfnamefont {Z.~L.}\ \bibnamefont {Dun}},
  \bibinfo {author} {\bibfnamefont {H.}~\bibnamefont {Zhou}}, \bibinfo {author}
  {\bibfnamefont {S.~A.}\ \bibnamefont {Grigera}},  \emph {et~al.},\
  }\href@noop {} {\bibfield  {journal} {\bibinfo  {journal} {Nature
  communications}\ }\textbf {\bibinfo {volume} {11}},\ \bibinfo {pages} {892}
  (\bibinfo {year} {2020})}\BibitemShut {NoStop}%
\bibitem [{\citenamefont {Ramirez}\ \emph {et~al.}(1999)\citenamefont
  {Ramirez}, \citenamefont {Hayashi}, \citenamefont {Cava}, \citenamefont
  {Siddharthan},\ and\ \citenamefont {Shastry}}]{ramirez1999zero}%
  \BibitemOpen
  \bibfield  {author} {\bibinfo {author} {\bibfnamefont {A.~P.}\ \bibnamefont
  {Ramirez}}, \bibinfo {author} {\bibfnamefont {A.}~\bibnamefont {Hayashi}},
  \bibinfo {author} {\bibfnamefont {R.~J.}\ \bibnamefont {Cava}}, \bibinfo
  {author} {\bibfnamefont {R.}~\bibnamefont {Siddharthan}}, \ and\ \bibinfo
  {author} {\bibfnamefont {B.}~\bibnamefont {Shastry}},\ }\href@noop {}
  {\bibfield  {journal} {\bibinfo  {journal} {Nature}\ }\textbf {\bibinfo
  {volume} {399}},\ \bibinfo {pages} {333} (\bibinfo {year}
  {1999})}\BibitemShut {NoStop}%
\bibitem [{\citenamefont {Castelnovo}\ \emph {et~al.}(2008)\citenamefont
  {Castelnovo}, \citenamefont {Moessner},\ and\ \citenamefont
  {Sondhi}}]{Castelnovo}%
  \BibitemOpen
  \bibfield  {author} {\bibinfo {author} {\bibfnamefont {C.}~\bibnamefont
  {Castelnovo}}, \bibinfo {author} {\bibfnamefont {R.}~\bibnamefont
  {Moessner}}, \ and\ \bibinfo {author} {\bibfnamefont {S.~L.}\ \bibnamefont
  {Sondhi}},\ }\href {\doibase 10.1038/nature06433} {\bibfield  {journal}
  {\bibinfo  {journal} {Nature}\ }\textbf {\bibinfo {volume} {451}},\ \bibinfo
  {pages} {42} (\bibinfo {year} {2008})}\BibitemShut {NoStop}%
\bibitem [{\citenamefont {Morris}\ \emph {et~al.}(2009)\citenamefont {Morris},
  \citenamefont {Tennant}, \citenamefont {Grigera}, \citenamefont {Klemke},
  \citenamefont {Castelnovo}, \citenamefont {Moessner}, \citenamefont
  {Czternasty}, \citenamefont {Meissner}, \citenamefont {Rule}, \citenamefont
  {Hoffmann} \emph {et~al.}}]{morris2009dirac}%
  \BibitemOpen
  \bibfield  {author} {\bibinfo {author} {\bibfnamefont {D.~J.~P.}\
  \bibnamefont {Morris}}, \bibinfo {author} {\bibfnamefont {D.}~\bibnamefont
  {Tennant}}, \bibinfo {author} {\bibfnamefont {S.~A.}\ \bibnamefont
  {Grigera}}, \bibinfo {author} {\bibfnamefont {B.}~\bibnamefont {Klemke}},
  \bibinfo {author} {\bibfnamefont {C.}~\bibnamefont {Castelnovo}}, \bibinfo
  {author} {\bibfnamefont {R.}~\bibnamefont {Moessner}}, \bibinfo {author}
  {\bibfnamefont {C.}~\bibnamefont {Czternasty}}, \bibinfo {author}
  {\bibfnamefont {M.}~\bibnamefont {Meissner}}, \bibinfo {author}
  {\bibfnamefont {K.}~\bibnamefont {Rule}}, \bibinfo {author} {\bibfnamefont
  {J.-U.}\ \bibnamefont {Hoffmann}},  \emph {et~al.},\ }\href@noop {}
  {\bibfield  {journal} {\bibinfo  {journal} {Science}\ }\textbf {\bibinfo
  {volume} {326}},\ \bibinfo {pages} {411} (\bibinfo {year}
  {2009})}\BibitemShut {NoStop}%
\bibitem [{\citenamefont {Guruciaga}\ \emph {et~al.}(2014)\citenamefont
  {Guruciaga}, \citenamefont {Grigera},\ and\ \citenamefont
  {Borzi}}]{Guruciaga14}%
  \BibitemOpen
  \bibfield  {author} {\bibinfo {author} {\bibfnamefont {P.~C.}\ \bibnamefont
  {Guruciaga}}, \bibinfo {author} {\bibfnamefont {S.~A.}\ \bibnamefont
  {Grigera}}, \ and\ \bibinfo {author} {\bibfnamefont {R.~A.}\ \bibnamefont
  {Borzi}},\ }\href {\doibase 10.1103/PhysRevB.90.184423} {\bibfield  {journal}
  {\bibinfo  {journal} {Phys. Rev. B}\ }\textbf {\bibinfo {volume} {90}},\
  \bibinfo {pages} {184423} (\bibinfo {year} {2014})}\BibitemShut {NoStop}%
\bibitem [{\citenamefont {Guruciaga}\ \emph {et~al.}(2016)\citenamefont
  {Guruciaga}, \citenamefont {Tarzia}, \citenamefont {Ferreyra}, \citenamefont
  {Cugliandolo}, \citenamefont {Grigera},\ and\ \citenamefont
  {Borzi}}]{Guruciaga16}%
  \BibitemOpen
  \bibfield  {author} {\bibinfo {author} {\bibfnamefont {P.~C.}\ \bibnamefont
  {Guruciaga}}, \bibinfo {author} {\bibfnamefont {M.}~\bibnamefont {Tarzia}},
  \bibinfo {author} {\bibfnamefont {M.~V.}\ \bibnamefont {Ferreyra}}, \bibinfo
  {author} {\bibfnamefont {L.~F.}\ \bibnamefont {Cugliandolo}}, \bibinfo
  {author} {\bibfnamefont {S.~A.}\ \bibnamefont {Grigera}}, \ and\ \bibinfo
  {author} {\bibfnamefont {R.~A.}\ \bibnamefont {Borzi}},\ }\href {\doibase
  10.1103/PhysRevLett.117.167203} {\bibfield  {journal} {\bibinfo  {journal}
  {Phys. Rev. Lett.}\ }\textbf {\bibinfo {volume} {117}},\ \bibinfo {pages}
  {167203} (\bibinfo {year} {2016})}\BibitemShut {NoStop}%
\bibitem [{\citenamefont {Pearce}\ \emph {et~al.}(2022)\citenamefont {Pearce},
  \citenamefont {G{\"o}tze}, \citenamefont {Szab{\'o}}, \citenamefont
  {Sikkenk}, \citenamefont {Lees}, \citenamefont {Boothroyd}, \citenamefont
  {Prabhakaran}, \citenamefont {Castelnovo},\ and\ \citenamefont
  {Goddard}}]{pearce2022monopoledensity}%
  \BibitemOpen
  \bibfield  {author} {\bibinfo {author} {\bibfnamefont {M.}~\bibnamefont
  {Pearce}}, \bibinfo {author} {\bibfnamefont {K.}~\bibnamefont {G{\"o}tze}},
  \bibinfo {author} {\bibfnamefont {A.}~\bibnamefont {Szab{\'o}}}, \bibinfo
  {author} {\bibfnamefont {T.}~\bibnamefont {Sikkenk}}, \bibinfo {author}
  {\bibfnamefont {M.}~\bibnamefont {Lees}}, \bibinfo {author} {\bibfnamefont
  {A.}~\bibnamefont {Boothroyd}}, \bibinfo {author} {\bibfnamefont
  {D.}~\bibnamefont {Prabhakaran}}, \bibinfo {author} {\bibfnamefont
  {C.}~\bibnamefont {Castelnovo}}, \ and\ \bibinfo {author} {\bibfnamefont
  {P.}~\bibnamefont {Goddard}},\ }\href@noop {} {\bibfield  {journal} {\bibinfo
   {journal} {Nature communications}\ }\textbf {\bibinfo {volume} {13}},\
  \bibinfo {pages} {1} (\bibinfo {year} {2022})}\BibitemShut {NoStop}%
\bibitem [{\citenamefont {Slobinsky}\ \emph {et~al.}(2019)\citenamefont
  {Slobinsky}, \citenamefont {Pili},\ and\ \citenamefont {Borzi}}]{pili_2019}%
  \BibitemOpen
  \bibfield  {author} {\bibinfo {author} {\bibfnamefont {D.}~\bibnamefont
  {Slobinsky}}, \bibinfo {author} {\bibfnamefont {L.}~\bibnamefont {Pili}}, \
  and\ \bibinfo {author} {\bibfnamefont {R.~A.}\ \bibnamefont {Borzi}},\ }\href
  {\doibase 10.1103/PhysRevB.100.020405} {\bibfield  {journal} {\bibinfo
  {journal} {Phys. Rev. B}\ }\textbf {\bibinfo {volume} {100}},\ \bibinfo
  {pages} {020405(R)} (\bibinfo {year} {2019})}\BibitemShut {NoStop}%
\bibitem [{\citenamefont {Jaubert}(2015)}]{jaubert2015holes}%
  \BibitemOpen
  \bibfield  {author} {\bibinfo {author} {\bibfnamefont {L.~D.~C.}\
  \bibnamefont {Jaubert}},\ }\href {\doibase 10.1142/S2010324715400056}
  {\bibfield  {journal} {\bibinfo  {journal} {SPIN}\ }\textbf {\bibinfo
  {volume} {05}},\ \bibinfo {pages} {1540005} (\bibinfo {year}
  {2015})}\BibitemShut {NoStop}%
\bibitem [{\citenamefont {Slobinsky}\ \emph {et~al.}(2021)\citenamefont
  {Slobinsky}, \citenamefont {Pili}, \citenamefont {Baglietto}, \citenamefont
  {Grigera},\ and\ \citenamefont {Borzi}}]{slobinsky2021monopole}%
  \BibitemOpen
  \bibfield  {author} {\bibinfo {author} {\bibfnamefont {D.}~\bibnamefont
  {Slobinsky}}, \bibinfo {author} {\bibfnamefont {L.}~\bibnamefont {Pili}},
  \bibinfo {author} {\bibfnamefont {G.}~\bibnamefont {Baglietto}}, \bibinfo
  {author} {\bibfnamefont {S.~A.}\ \bibnamefont {Grigera}}, \ and\ \bibinfo
  {author} {\bibfnamefont {R.~A.}\ \bibnamefont {Borzi}},\ }\href@noop {}
  {\bibfield  {journal} {\bibinfo  {journal} {Communications Physics}\ }\textbf
  {\bibinfo {volume} {4}},\ \bibinfo {pages} {1} (\bibinfo {year}
  {2021})}\BibitemShut {NoStop}%
\bibitem [{\citenamefont {Khomskii}(2012)}]{Khomskii2012electric}%
  \BibitemOpen
  \bibfield  {author} {\bibinfo {author} {\bibfnamefont {D.~I.}\ \bibnamefont
  {Khomskii}},\ }\href {https://www.nature.com/articles/ncomms1904} {\bibfield
  {journal} {\bibinfo  {journal} {Nat. Commun.}\ }\textbf {\bibinfo {volume}
  {3}},\ \bibinfo {pages} {904} (\bibinfo {year} {2012})}\BibitemShut {NoStop}%
\bibitem [{\citenamefont {Grams}\ \emph {et~al.}(2014)\citenamefont {Grams},
  \citenamefont {Valldor}, \citenamefont {Garst},\ and\ \citenamefont
  {Hemberger}}]{grams2014critical}%
  \BibitemOpen
  \bibfield  {author} {\bibinfo {author} {\bibfnamefont {C.~P.}\ \bibnamefont
  {Grams}}, \bibinfo {author} {\bibfnamefont {M.}~\bibnamefont {Valldor}},
  \bibinfo {author} {\bibfnamefont {M.}~\bibnamefont {Garst}}, \ and\ \bibinfo
  {author} {\bibfnamefont {J.}~\bibnamefont {Hemberger}},\ }\href@noop {}
  {\bibfield  {journal} {\bibinfo  {journal} {Nature communications}\ }\textbf
  {\bibinfo {volume} {5}},\ \bibinfo {pages} {4853} (\bibinfo {year}
  {2014})}\BibitemShut {NoStop}%
\bibitem [{\citenamefont {Jin}\ \emph {et~al.}(2020)\citenamefont {Jin},
  \citenamefont {Liu}, \citenamefont {Chang}, \citenamefont {Zhang},
  \citenamefont {Wang}, \citenamefont {Liu}, \citenamefont {Wang},
  \citenamefont {Sun}, \citenamefont {Chen}, \citenamefont {Sun} \emph
  {et~al.}}]{jin2020experimental}%
  \BibitemOpen
  \bibfield  {author} {\bibinfo {author} {\bibfnamefont {F.}~\bibnamefont
  {Jin}}, \bibinfo {author} {\bibfnamefont {C.}~\bibnamefont {Liu}}, \bibinfo
  {author} {\bibfnamefont {Y.}~\bibnamefont {Chang}}, \bibinfo {author}
  {\bibfnamefont {A.}~\bibnamefont {Zhang}}, \bibinfo {author} {\bibfnamefont
  {Y.}~\bibnamefont {Wang}}, \bibinfo {author} {\bibfnamefont {W.}~\bibnamefont
  {Liu}}, \bibinfo {author} {\bibfnamefont {X.}~\bibnamefont {Wang}}, \bibinfo
  {author} {\bibfnamefont {Y.}~\bibnamefont {Sun}}, \bibinfo {author}
  {\bibfnamefont {G.}~\bibnamefont {Chen}}, \bibinfo {author} {\bibfnamefont
  {X.}~\bibnamefont {Sun}},  \emph {et~al.},\ }\href@noop {} {\bibfield
  {journal} {\bibinfo  {journal} {Physical review letters}\ }\textbf {\bibinfo
  {volume} {124}},\ \bibinfo {pages} {087601} (\bibinfo {year}
  {2020})}\BibitemShut {NoStop}%
\bibitem [{\citenamefont {Jaubert}\ and\ \citenamefont
  {Moessner}(2015)}]{Jaubert2015crystallog}%
  \BibitemOpen
  \bibfield  {author} {\bibinfo {author} {\bibfnamefont {L.~D.~C.}\
  \bibnamefont {Jaubert}}\ and\ \bibinfo {author} {\bibfnamefont
  {R.}~\bibnamefont {Moessner}},\ }\href {\doibase 10.1103/PhysRevB.91.214422}
  {\bibfield  {journal} {\bibinfo  {journal} {Phys. Rev. B}\ }\textbf {\bibinfo
  {volume} {91}},\ \bibinfo {pages} {214422} (\bibinfo {year}
  {2015})}\BibitemShut {NoStop}%
\bibitem [{\citenamefont {Katsufuji}\ and\ \citenamefont
  {Takagi}(2004)}]{Katsufuji2004magnetocapacitance}%
  \BibitemOpen
  \bibfield  {author} {\bibinfo {author} {\bibfnamefont {T.}~\bibnamefont
  {Katsufuji}}\ and\ \bibinfo {author} {\bibfnamefont {H.}~\bibnamefont
  {Takagi}},\ }\href {\doibase 10.1103/PhysRevB.69.064422} {\bibfield
  {journal} {\bibinfo  {journal} {Phys. Rev. B}\ }\textbf {\bibinfo {volume}
  {69}},\ \bibinfo {pages} {064422} (\bibinfo {year} {2004})}\BibitemShut
  {NoStop}%
\bibitem [{\citenamefont {Saito}\ \emph {et~al.}(2005)\citenamefont {Saito},
  \citenamefont {Higashinaka},\ and\ \citenamefont
  {Maeno}}]{Saito2005magnetodielectric}%
  \BibitemOpen
  \bibfield  {author} {\bibinfo {author} {\bibfnamefont {M.}~\bibnamefont
  {Saito}}, \bibinfo {author} {\bibfnamefont {R.}~\bibnamefont {Higashinaka}},
  \ and\ \bibinfo {author} {\bibfnamefont {Y.}~\bibnamefont {Maeno}},\ }\href
  {\doibase 10.1103/PhysRevB.72.144422} {\bibfield  {journal} {\bibinfo
  {journal} {Phys. Rev. B}\ }\textbf {\bibinfo {volume} {72}},\ \bibinfo
  {pages} {144422} (\bibinfo {year} {2005})}\BibitemShut {NoStop}%
\bibitem [{\citenamefont {Liu}\ \emph {et~al.}(2013)\citenamefont {Liu},
  \citenamefont {Lin}, \citenamefont {Liu}, \citenamefont {Yan}, \citenamefont
  {Dong},\ and\ \citenamefont {Liu}}]{liu2013multiferroicity}%
  \BibitemOpen
  \bibfield  {author} {\bibinfo {author} {\bibfnamefont {D.}~\bibnamefont
  {Liu}}, \bibinfo {author} {\bibfnamefont {L.}~\bibnamefont {Lin}}, \bibinfo
  {author} {\bibfnamefont {M.}~\bibnamefont {Liu}}, \bibinfo {author}
  {\bibfnamefont {Z.}~\bibnamefont {Yan}}, \bibinfo {author} {\bibfnamefont
  {S.}~\bibnamefont {Dong}}, \ and\ \bibinfo {author} {\bibfnamefont {J.-M.}\
  \bibnamefont {Liu}},\ }\href@noop {} {\bibfield  {journal} {\bibinfo
  {journal} {Journal of Applied Physics}\ }\textbf {\bibinfo {volume} {113}}
  (\bibinfo {year} {2013})}\BibitemShut {NoStop}%
\bibitem [{\citenamefont {Lin}\ \emph {et~al.}(2015)\citenamefont {Lin},
  \citenamefont {Xie}, \citenamefont {Wen}, \citenamefont {Dong}, \citenamefont
  {Yan},\ and\ \citenamefont {Liu}}]{lin2015experimental}%
  \BibitemOpen
  \bibfield  {author} {\bibinfo {author} {\bibfnamefont {L.}~\bibnamefont
  {Lin}}, \bibinfo {author} {\bibfnamefont {Y.}~\bibnamefont {Xie}}, \bibinfo
  {author} {\bibfnamefont {J.}~\bibnamefont {Wen}}, \bibinfo {author}
  {\bibfnamefont {S.}~\bibnamefont {Dong}}, \bibinfo {author} {\bibfnamefont
  {Z.}~\bibnamefont {Yan}}, \ and\ \bibinfo {author} {\bibfnamefont
  {J.}~\bibnamefont {Liu}},\ }\href@noop {} {\bibfield  {journal} {\bibinfo
  {journal} {New Journal of Physics}\ }\textbf {\bibinfo {volume} {17}},\
  \bibinfo {pages} {123018} (\bibinfo {year} {2015})}\BibitemShut {NoStop}%
\bibitem [{\citenamefont {Yadav}\ and\ \citenamefont
  {Upadhyay}(2019)}]{yadav2019magnetodielectric}%
  \BibitemOpen
  \bibfield  {author} {\bibinfo {author} {\bibfnamefont {P.~K.}\ \bibnamefont
  {Yadav}}\ and\ \bibinfo {author} {\bibfnamefont {C.}~\bibnamefont
  {Upadhyay}},\ }\href@noop {} {\bibfield  {journal} {\bibinfo  {journal}
  {Journal of Superconductivity and Novel Magnetism}\ }\textbf {\bibinfo
  {volume} {32}},\ \bibinfo {pages} {2267} (\bibinfo {year}
  {2019})}\BibitemShut {NoStop}%
\bibitem [{\citenamefont {Khomskii}(2021)}]{khomskii2021electric}%
  \BibitemOpen
  \bibfield  {author} {\bibinfo {author} {\bibfnamefont {D.}~\bibnamefont
  {Khomskii}},\ }\href@noop {} {\bibfield  {journal} {\bibinfo  {journal}
  {Nature Communications}\ }\textbf {\bibinfo {volume} {12}},\ \bibinfo {pages}
  {3047} (\bibinfo {year} {2021})}\BibitemShut {NoStop}%
\bibitem [{\citenamefont {Opherden}\ \emph {et~al.}(2017)\citenamefont
  {Opherden}, \citenamefont {Hornung}, \citenamefont {Herrmannsd{\"o}rfer},
  \citenamefont {Xu}, \citenamefont {Islam}, \citenamefont {Lake},\ and\
  \citenamefont {Wosnitza}}]{opherden2017evolution}%
  \BibitemOpen
  \bibfield  {author} {\bibinfo {author} {\bibfnamefont {L.}~\bibnamefont
  {Opherden}}, \bibinfo {author} {\bibfnamefont {J.}~\bibnamefont {Hornung}},
  \bibinfo {author} {\bibfnamefont {T.}~\bibnamefont {Herrmannsd{\"o}rfer}},
  \bibinfo {author} {\bibfnamefont {J.}~\bibnamefont {Xu}}, \bibinfo {author}
  {\bibfnamefont {A.~T. M.~N.}\ \bibnamefont {Islam}}, \bibinfo {author}
  {\bibfnamefont {B.}~\bibnamefont {Lake}}, \ and\ \bibinfo {author}
  {\bibfnamefont {J.}~\bibnamefont {Wosnitza}},\ }\href@noop {} {\bibfield
  {journal} {\bibinfo  {journal} {Physical Review B}\ }\textbf {\bibinfo
  {volume} {95}},\ \bibinfo {pages} {184418} (\bibinfo {year}
  {2017})}\BibitemShut {NoStop}%
\bibitem [{\citenamefont {Onoda}\ and\ \citenamefont
  {Tanaka}(2011)}]{Onoda2011exchange}%
  \BibitemOpen
  \bibfield  {author} {\bibinfo {author} {\bibfnamefont {S.}~\bibnamefont
  {Onoda}}\ and\ \bibinfo {author} {\bibfnamefont {Y.}~\bibnamefont {Tanaka}},\
  }\href@noop {} {\bibfield  {journal} {\bibinfo  {journal} {Physical Review
  B}\ }\textbf {\bibinfo {volume} {83}},\ \bibinfo {pages} {094411} (\bibinfo
  {year} {2011})}\BibitemShut {NoStop}%
\bibitem [{\citenamefont {Tomasello}\ \emph {et~al.}(2018)\citenamefont
  {Tomasello}, \citenamefont {Castelnovo}, \citenamefont {Moessner},\ and\
  \citenamefont {Quintanilla}}]{Tomasello2018correlated}%
  \BibitemOpen
  \bibfield  {author} {\bibinfo {author} {\bibfnamefont {B.}~\bibnamefont
  {Tomasello}}, \bibinfo {author} {\bibfnamefont {C.}~\bibnamefont
  {Castelnovo}}, \bibinfo {author} {\bibfnamefont {R.}~\bibnamefont
  {Moessner}}, \ and\ \bibinfo {author} {\bibfnamefont {J.}~\bibnamefont
  {Quintanilla}},\ }\href@noop {} {\bibfield  {journal} {\bibinfo  {journal}
  {arXiv preprint arXiv:1810.11469}\ } (\bibinfo {year} {2018})}\BibitemShut
  {NoStop}%
\bibitem [{\citenamefont {Sazonov}\ \emph {et~al.}(2013)\citenamefont
  {Sazonov}, \citenamefont {Gukasov}, \citenamefont {Cao}, \citenamefont
  {Bonville}, \citenamefont {Ressouche}, \citenamefont {Decorse},\ and\
  \citenamefont {Mirebeau}}]{Sazonov2013}%
  \BibitemOpen
  \bibfield  {author} {\bibinfo {author} {\bibfnamefont {A.~P.}\ \bibnamefont
  {Sazonov}}, \bibinfo {author} {\bibfnamefont {A.}~\bibnamefont {Gukasov}},
  \bibinfo {author} {\bibfnamefont {H.~B.}\ \bibnamefont {Cao}}, \bibinfo
  {author} {\bibfnamefont {P.}~\bibnamefont {Bonville}}, \bibinfo {author}
  {\bibfnamefont {E.}~\bibnamefont {Ressouche}}, \bibinfo {author}
  {\bibfnamefont {C.}~\bibnamefont {Decorse}}, \ and\ \bibinfo {author}
  {\bibfnamefont {I.}~\bibnamefont {Mirebeau}},\ }\href@noop {} {\bibfield
  {journal} {\bibinfo  {journal} {Physical Review B}\ }\textbf {\bibinfo
  {volume} {88}},\ \bibinfo {pages} {184428} (\bibinfo {year}
  {2013})}\BibitemShut {NoStop}%
\bibitem [{\citenamefont {Slobinsky}\ \emph {et~al.}(2018)\citenamefont
  {Slobinsky}, \citenamefont {Baglietto},\ and\ \citenamefont
  {Borzi}}]{Slobinsky2018charge}%
  \BibitemOpen
  \bibfield  {author} {\bibinfo {author} {\bibfnamefont {D.}~\bibnamefont
  {Slobinsky}}, \bibinfo {author} {\bibfnamefont {G.}~\bibnamefont
  {Baglietto}}, \ and\ \bibinfo {author} {\bibfnamefont {R.~A.}\ \bibnamefont
  {Borzi}},\ }\href {\doibase 10.1103/PhysRevB.97.174422} {\bibfield  {journal}
  {\bibinfo  {journal} {Phys. Rev. B}\ }\textbf {\bibinfo {volume} {97}},\
  \bibinfo {pages} {174422} (\bibinfo {year} {2018})}\BibitemShut {NoStop}%
\bibitem [{\citenamefont {Vignau}\ and\ \citenamefont
  {Borzi}(2023)}]{Vignau23}%
  \BibitemOpen
  \bibfield  {author} {\bibinfo {author} {\bibfnamefont {T.}~\bibnamefont
  {Vignau}}\ and\ \bibinfo {author} {\bibfnamefont {R.~A.}\ \bibnamefont
  {Borzi}},\ }\href@noop {} {\bibfield  {journal} {\bibinfo  {journal} {in
  preparation}\ } (\bibinfo {year} {2023})}\BibitemShut {NoStop}%
\bibitem [{\citenamefont {Fennell}\ \emph {et~al.}(2002)\citenamefont
  {Fennell}, \citenamefont {Petrenko}, \citenamefont {Balakrishnan},
  \citenamefont {Bramwell}, \citenamefont {Champion}, \citenamefont {F{\aa}k},
  \citenamefont {Harris},\ and\ \citenamefont {Paul}}]{fennell2002field}%
  \BibitemOpen
  \bibfield  {author} {\bibinfo {author} {\bibfnamefont {T.}~\bibnamefont
  {Fennell}}, \bibinfo {author} {\bibfnamefont {O.}~\bibnamefont {Petrenko}},
  \bibinfo {author} {\bibfnamefont {G.}~\bibnamefont {Balakrishnan}}, \bibinfo
  {author} {\bibfnamefont {S.}~\bibnamefont {Bramwell}}, \bibinfo {author}
  {\bibfnamefont {J.}~\bibnamefont {Champion}}, \bibinfo {author}
  {\bibfnamefont {B.}~\bibnamefont {F{\aa}k}}, \bibinfo {author} {\bibfnamefont
  {M.}~\bibnamefont {Harris}}, \ and\ \bibinfo {author} {\bibfnamefont {D.~M.}\
  \bibnamefont {Paul}},\ }\href@noop {} {\bibfield  {journal} {\bibinfo
  {journal} {Applied Physics A}\ }\textbf {\bibinfo {volume} {74}},\ \bibinfo
  {pages} {s889} (\bibinfo {year} {2002})}\BibitemShut {NoStop}%
\bibitem [{\citenamefont {Sakakibara}\ \emph {et~al.}(2003)\citenamefont
  {Sakakibara}, \citenamefont {Tayama}, \citenamefont {Hiroi}, \citenamefont
  {Matsuhira},\ and\ \citenamefont {Takagi}}]{Sakakibara03}%
  \BibitemOpen
  \bibfield  {author} {\bibinfo {author} {\bibfnamefont {T.}~\bibnamefont
  {Sakakibara}}, \bibinfo {author} {\bibfnamefont {T.}~\bibnamefont {Tayama}},
  \bibinfo {author} {\bibfnamefont {Z.}~\bibnamefont {Hiroi}}, \bibinfo
  {author} {\bibfnamefont {K.}~\bibnamefont {Matsuhira}}, \ and\ \bibinfo
  {author} {\bibfnamefont {S.}~\bibnamefont {Takagi}},\ }\href {\doibase
  10.1103/PhysRevLett.90.207205} {\bibfield  {journal} {\bibinfo  {journal}
  {Phys. Rev. Lett.}\ }\textbf {\bibinfo {volume} {90}},\ \bibinfo {pages}
  {207205} (\bibinfo {year} {2003})}\BibitemShut {NoStop}%
\bibitem [{\citenamefont {Isakov}\ \emph {et~al.}(2004)\citenamefont {Isakov},
  \citenamefont {Raman}, \citenamefont {Moessner},\ and\ \citenamefont
  {Sondhi}}]{isakov2004magnetization}%
  \BibitemOpen
  \bibfield  {author} {\bibinfo {author} {\bibfnamefont {S.~V.}\ \bibnamefont
  {Isakov}}, \bibinfo {author} {\bibfnamefont {K.~S.}\ \bibnamefont {Raman}},
  \bibinfo {author} {\bibfnamefont {R.}~\bibnamefont {Moessner}}, \ and\
  \bibinfo {author} {\bibfnamefont {S.~L.}\ \bibnamefont {Sondhi}},\
  }\href@noop {} {\bibfield  {journal} {\bibinfo  {journal} {Physical Review
  B}\ }\textbf {\bibinfo {volume} {70}},\ \bibinfo {pages} {104418} (\bibinfo
  {year} {2004})}\BibitemShut {NoStop}%
\bibitem [{\citenamefont {Molavian}\ and\ \citenamefont
  {Gingras}(2009)}]{molavian2009proposal}%
  \BibitemOpen
  \bibfield  {author} {\bibinfo {author} {\bibfnamefont {H.~R.}\ \bibnamefont
  {Molavian}}\ and\ \bibinfo {author} {\bibfnamefont {M.~J.}\ \bibnamefont
  {Gingras}},\ }\href@noop {} {\bibfield  {journal} {\bibinfo  {journal}
  {Journal of Physics: Condensed Matter}\ }\textbf {\bibinfo {volume} {21}},\
  \bibinfo {pages} {172201} (\bibinfo {year} {2009})}\BibitemShut {NoStop}%
\bibitem [{\citenamefont {Lhotel}\ \emph {et~al.}(2015)\citenamefont {Lhotel},
  \citenamefont {Petit}, \citenamefont {Guitteny}, \citenamefont {Florea},
  \citenamefont {Ciomaga~Hatnean}, \citenamefont {Colin}, \citenamefont
  {Ressouche}, \citenamefont {Lees},\ and\ \citenamefont
  {Balakrishnan}}]{lhotel2015fluctuations}%
  \BibitemOpen
  \bibfield  {author} {\bibinfo {author} {\bibfnamefont {E.}~\bibnamefont
  {Lhotel}}, \bibinfo {author} {\bibfnamefont {S.}~\bibnamefont {Petit}},
  \bibinfo {author} {\bibfnamefont {S.}~\bibnamefont {Guitteny}}, \bibinfo
  {author} {\bibfnamefont {O.}~\bibnamefont {Florea}}, \bibinfo {author}
  {\bibfnamefont {M.}~\bibnamefont {Ciomaga~Hatnean}}, \bibinfo {author}
  {\bibfnamefont {C.}~\bibnamefont {Colin}}, \bibinfo {author} {\bibfnamefont
  {E.}~\bibnamefont {Ressouche}}, \bibinfo {author} {\bibfnamefont {M.~R.}\
  \bibnamefont {Lees}}, \ and\ \bibinfo {author} {\bibfnamefont
  {G.}~\bibnamefont {Balakrishnan}},\ }\href@noop {} {\bibfield  {journal}
  {\bibinfo  {journal} {Physical Review Letters}\ }\textbf {\bibinfo {volume}
  {115}},\ \bibinfo {pages} {197202} (\bibinfo {year} {2015})}\BibitemShut
  {NoStop}%
\bibitem [{\citenamefont {Tian}\ \emph {et~al.}(2016)\citenamefont {Tian},
  \citenamefont {Kohama}, \citenamefont {Tomita}, \citenamefont {Ishizuka},
  \citenamefont {Hsieh}, \citenamefont {Ishikawa}, \citenamefont {Kindo},
  \citenamefont {Balents},\ and\ \citenamefont {Nakatsuji}}]{tian2016field}%
  \BibitemOpen
  \bibfield  {author} {\bibinfo {author} {\bibfnamefont {Z.}~\bibnamefont
  {Tian}}, \bibinfo {author} {\bibfnamefont {Y.}~\bibnamefont {Kohama}},
  \bibinfo {author} {\bibfnamefont {T.}~\bibnamefont {Tomita}}, \bibinfo
  {author} {\bibfnamefont {H.}~\bibnamefont {Ishizuka}}, \bibinfo {author}
  {\bibfnamefont {T.~H.}\ \bibnamefont {Hsieh}}, \bibinfo {author}
  {\bibfnamefont {J.~J.}\ \bibnamefont {Ishikawa}}, \bibinfo {author}
  {\bibfnamefont {K.}~\bibnamefont {Kindo}}, \bibinfo {author} {\bibfnamefont
  {L.}~\bibnamefont {Balents}}, \ and\ \bibinfo {author} {\bibfnamefont
  {S.}~\bibnamefont {Nakatsuji}},\ }\href@noop {} {\bibfield  {journal}
  {\bibinfo  {journal} {Nature Physics}\ }\textbf {\bibinfo {volume} {12}},\
  \bibinfo {pages} {134} (\bibinfo {year} {2016})}\BibitemShut {NoStop}%
\bibitem [{\citenamefont {Opherden}\ \emph {et~al.}(2018)\citenamefont
  {Opherden}, \citenamefont {Bilitewski}, \citenamefont {Hornung},
  \citenamefont {Herrmannsd{\"o}rfer}, \citenamefont {Samartzis}, \citenamefont
  {Islam}, \citenamefont {Anand}, \citenamefont {Lake}, \citenamefont
  {Moessner},\ and\ \citenamefont {Wosnitza}}]{opherden2018inverted}%
  \BibitemOpen
  \bibfield  {author} {\bibinfo {author} {\bibfnamefont {L.}~\bibnamefont
  {Opherden}}, \bibinfo {author} {\bibfnamefont {T.}~\bibnamefont
  {Bilitewski}}, \bibinfo {author} {\bibfnamefont {J.}~\bibnamefont {Hornung}},
  \bibinfo {author} {\bibfnamefont {T.}~\bibnamefont {Herrmannsd{\"o}rfer}},
  \bibinfo {author} {\bibfnamefont {A.}~\bibnamefont {Samartzis}}, \bibinfo
  {author} {\bibfnamefont {A.~T. M.~N.}\ \bibnamefont {Islam}}, \bibinfo
  {author} {\bibfnamefont {V.~K.}\ \bibnamefont {Anand}}, \bibinfo {author}
  {\bibfnamefont {B.}~\bibnamefont {Lake}}, \bibinfo {author} {\bibfnamefont
  {R.}~\bibnamefont {Moessner}}, \ and\ \bibinfo {author} {\bibfnamefont
  {J.}~\bibnamefont {Wosnitza}},\ }\href@noop {} {\bibfield  {journal}
  {\bibinfo  {journal} {Physical Review B}\ }\textbf {\bibinfo {volume} {98}},\
  \bibinfo {pages} {180403(R)} (\bibinfo {year} {2018})}\BibitemShut {NoStop}%
\bibitem [{\citenamefont {Xu}\ \emph {et~al.}(2019)\citenamefont {Xu},
  \citenamefont {Benton}, \citenamefont {Anand}, \citenamefont {Islam},
  \citenamefont {Guidi}, \citenamefont {Ehlers}, \citenamefont {Feng},
  \citenamefont {Su}, \citenamefont {Sakai}, \citenamefont {Gegenwart},\ and\
  \citenamefont {Lake}}]{xu2019anisotropic}%
  \BibitemOpen
  \bibfield  {author} {\bibinfo {author} {\bibfnamefont {J.}~\bibnamefont
  {Xu}}, \bibinfo {author} {\bibfnamefont {O.}~\bibnamefont {Benton}}, \bibinfo
  {author} {\bibfnamefont {V.~K.}\ \bibnamefont {Anand}}, \bibinfo {author}
  {\bibfnamefont {A.~T. M.~N.}\ \bibnamefont {Islam}}, \bibinfo {author}
  {\bibfnamefont {T.}~\bibnamefont {Guidi}}, \bibinfo {author} {\bibfnamefont
  {G.}~\bibnamefont {Ehlers}}, \bibinfo {author} {\bibfnamefont
  {E.}~\bibnamefont {Feng}}, \bibinfo {author} {\bibfnamefont {Y.}~\bibnamefont
  {Su}}, \bibinfo {author} {\bibfnamefont {A.}~\bibnamefont {Sakai}}, \bibinfo
  {author} {\bibfnamefont {P.}~\bibnamefont {Gegenwart}}, \ and\ \bibinfo
  {author} {\bibfnamefont {B.}~\bibnamefont {Lake}},\ }\href@noop {} {\bibfield
   {journal} {\bibinfo  {journal} {Physical Review B}\ }\textbf {\bibinfo
  {volume} {99}},\ \bibinfo {pages} {144420} (\bibinfo {year}
  {2019})}\BibitemShut {NoStop}%
\bibitem [{\citenamefont {Guruciaga}\ and\ \citenamefont
  {Borzi}(2019)}]{guruciaga2019monte}%
  \BibitemOpen
  \bibfield  {author} {\bibinfo {author} {\bibfnamefont {P.~C.}\ \bibnamefont
  {Guruciaga}}\ and\ \bibinfo {author} {\bibfnamefont {R.~A.}\ \bibnamefont
  {Borzi}},\ }\href@noop {} {\bibfield  {journal} {\bibinfo  {journal}
  {Physical Review B}\ }\textbf {\bibinfo {volume} {100}},\ \bibinfo {pages}
  {174404} (\bibinfo {year} {2019})}\BibitemShut {NoStop}%
\bibitem [{\citenamefont {Fennell}\ \emph {et~al.}(2014)\citenamefont
  {Fennell}, \citenamefont {Kenzelmann}, \citenamefont {Roessli}, \citenamefont
  {Mutka}, \citenamefont {Ollivier}, \citenamefont {Ruminy}, \citenamefont
  {Stuhr}, \citenamefont {Zaharko}, \citenamefont {Bovo}, \citenamefont
  {Cervellino}, \citenamefont {Haas},\ and\ \citenamefont
  {Cava}}]{Fennell2014magnetoelastic}%
  \BibitemOpen
  \bibfield  {author} {\bibinfo {author} {\bibfnamefont {T.}~\bibnamefont
  {Fennell}}, \bibinfo {author} {\bibfnamefont {M.}~\bibnamefont {Kenzelmann}},
  \bibinfo {author} {\bibfnamefont {B.}~\bibnamefont {Roessli}}, \bibinfo
  {author} {\bibfnamefont {H.}~\bibnamefont {Mutka}}, \bibinfo {author}
  {\bibfnamefont {J.}~\bibnamefont {Ollivier}}, \bibinfo {author}
  {\bibfnamefont {M.}~\bibnamefont {Ruminy}}, \bibinfo {author} {\bibfnamefont
  {U.}~\bibnamefont {Stuhr}}, \bibinfo {author} {\bibfnamefont
  {O.}~\bibnamefont {Zaharko}}, \bibinfo {author} {\bibfnamefont
  {L.}~\bibnamefont {Bovo}}, \bibinfo {author} {\bibfnamefont {A.}~\bibnamefont
  {Cervellino}}, \bibinfo {author} {\bibfnamefont {M.~K.}\ \bibnamefont
  {Haas}}, \ and\ \bibinfo {author} {\bibfnamefont {R.~J.}\ \bibnamefont
  {Cava}},\ }\href {\doibase 10.1103/PhysRevLett.112.017203} {\bibfield
  {journal} {\bibinfo  {journal} {Phys. Rev. Lett.}\ }\textbf {\bibinfo
  {volume} {112}},\ \bibinfo {pages} {017203} (\bibinfo {year}
  {2014})}\BibitemShut {NoStop}%
\bibitem [{\citenamefont {Ruff}\ \emph {et~al.}(2010)\citenamefont {Ruff},
  \citenamefont {Islam}, \citenamefont {Clancy}, \citenamefont {Ross},
  \citenamefont {Nojiri}, \citenamefont {Matsuda}, \citenamefont {Dabkowska},
  \citenamefont {Dabkowski},\ and\ \citenamefont
  {Gaulin}}]{ruff2010magnetoelastics}%
  \BibitemOpen
  \bibfield  {author} {\bibinfo {author} {\bibfnamefont {J.~P.~C.}\
  \bibnamefont {Ruff}}, \bibinfo {author} {\bibfnamefont {Z.}~\bibnamefont
  {Islam}}, \bibinfo {author} {\bibfnamefont {J.~P.}\ \bibnamefont {Clancy}},
  \bibinfo {author} {\bibfnamefont {K.~A.}\ \bibnamefont {Ross}}, \bibinfo
  {author} {\bibfnamefont {H.}~\bibnamefont {Nojiri}}, \bibinfo {author}
  {\bibfnamefont {Y.~H.}\ \bibnamefont {Matsuda}}, \bibinfo {author}
  {\bibfnamefont {H.~A.}\ \bibnamefont {Dabkowska}}, \bibinfo {author}
  {\bibfnamefont {A.~D.}\ \bibnamefont {Dabkowski}}, \ and\ \bibinfo {author}
  {\bibfnamefont {B.~D.}\ \bibnamefont {Gaulin}},\ }\href {\doibase
  10.1103/PhysRevLett.105.077203} {\bibfield  {journal} {\bibinfo  {journal}
  {Phys. Rev. Lett.}\ }\textbf {\bibinfo {volume} {105}},\ \bibinfo {pages}
  {077203} (\bibinfo {year} {2010})}\BibitemShut {NoStop}%
\bibitem [{\citenamefont {Edberg}\ \emph {et~al.}(2019)\citenamefont {Edberg},
  \citenamefont {Sandberg}, \citenamefont {Bakke}, \citenamefont {Haubro},
  \citenamefont {Folkers}, \citenamefont {Mangin-Thro}, \citenamefont {Wildes},
  \citenamefont {Zaharko}, \citenamefont {Guthrie}, \citenamefont {Holmes}
  \emph {et~al.}}]{edberg2019dipolar}%
  \BibitemOpen
  \bibfield  {author} {\bibinfo {author} {\bibfnamefont {R.}~\bibnamefont
  {Edberg}}, \bibinfo {author} {\bibfnamefont {L.~O.}\ \bibnamefont
  {Sandberg}}, \bibinfo {author} {\bibfnamefont {I.~M.~B.}\ \bibnamefont
  {Bakke}}, \bibinfo {author} {\bibfnamefont {M.~L.}\ \bibnamefont {Haubro}},
  \bibinfo {author} {\bibfnamefont {L.~C.}\ \bibnamefont {Folkers}}, \bibinfo
  {author} {\bibfnamefont {L.}~\bibnamefont {Mangin-Thro}}, \bibinfo {author}
  {\bibfnamefont {A.}~\bibnamefont {Wildes}}, \bibinfo {author} {\bibfnamefont
  {O.}~\bibnamefont {Zaharko}}, \bibinfo {author} {\bibfnamefont
  {M.}~\bibnamefont {Guthrie}}, \bibinfo {author} {\bibfnamefont {A.~T.}\
  \bibnamefont {Holmes}},  \emph {et~al.},\ }\href@noop {} {\bibfield
  {journal} {\bibinfo  {journal} {Physical Review B}\ }\textbf {\bibinfo
  {volume} {100}},\ \bibinfo {pages} {144436} (\bibinfo {year}
  {2019})}\BibitemShut {NoStop}%
\bibitem [{\citenamefont {Gupta}\ \emph {et~al.}(2009)\citenamefont {Gupta},
  \citenamefont {Singh}, \citenamefont {Kumar}, \citenamefont {Rani} \emph
  {et~al.}}]{gupta2009lattice}%
  \BibitemOpen
  \bibfield  {author} {\bibinfo {author} {\bibfnamefont {H.}~\bibnamefont
  {Gupta}}, \bibinfo {author} {\bibfnamefont {J.}~\bibnamefont {Singh}},
  \bibinfo {author} {\bibfnamefont {S.}~\bibnamefont {Kumar}}, \bibinfo
  {author} {\bibfnamefont {N.}~\bibnamefont {Rani}},  \emph {et~al.},\
  }\href@noop {} {\bibfield  {journal} {\bibinfo  {journal} {Journal of
  Molecular Structure}\ }\textbf {\bibinfo {volume} {937}},\ \bibinfo {pages}
  {136} (\bibinfo {year} {2009})}\BibitemShut {NoStop}%
\bibitem [{\citenamefont {Kushwaha}(2017)}]{kushwaha2017vibrational}%
  \BibitemOpen
  \bibfield  {author} {\bibinfo {author} {\bibfnamefont {A.}~\bibnamefont
  {Kushwaha}},\ }\href@noop {} {\bibfield  {journal} {\bibinfo  {journal}
  {International Journal of Modern Physics B}\ }\textbf {\bibinfo {volume}
  {31}},\ \bibinfo {pages} {1750145} (\bibinfo {year} {2017})}\BibitemShut
  {NoStop}%
\bibitem [{\citenamefont {Sarkar}\ and\ \citenamefont
  {Mukhopadhyay}(2014)}]{sarkar2014dynamics}%
  \BibitemOpen
  \bibfield  {author} {\bibinfo {author} {\bibfnamefont {A.}~\bibnamefont
  {Sarkar}}\ and\ \bibinfo {author} {\bibfnamefont {S.}~\bibnamefont
  {Mukhopadhyay}},\ }\href@noop {} {\bibfield  {journal} {\bibinfo  {journal}
  {Physical Review B}\ }\textbf {\bibinfo {volume} {90}},\ \bibinfo {pages}
  {165129} (\bibinfo {year} {2014})}\BibitemShut {NoStop}%
\bibitem [{\citenamefont {Katsura}\ \emph {et~al.}(2005)\citenamefont
  {Katsura}, \citenamefont {Nagaosa},\ and\ \citenamefont
  {Balatsky}}]{katsura2005spin}%
  \BibitemOpen
  \bibfield  {author} {\bibinfo {author} {\bibfnamefont {H.}~\bibnamefont
  {Katsura}}, \bibinfo {author} {\bibfnamefont {N.}~\bibnamefont {Nagaosa}}, \
  and\ \bibinfo {author} {\bibfnamefont {A.~V.}\ \bibnamefont {Balatsky}},\
  }\href@noop {} {\bibfield  {journal} {\bibinfo  {journal} {Physical review
  letters}\ }\textbf {\bibinfo {volume} {95}},\ \bibinfo {pages} {057205}
  (\bibinfo {year} {2005})}\BibitemShut {NoStop}%
\bibitem [{\citenamefont {Krey}\ \emph {et~al.}(2012)\citenamefont {Krey},
  \citenamefont {Legl}, \citenamefont {Dunsiger}, \citenamefont {Meven},
  \citenamefont {Gardner}, \citenamefont {Roper},\ and\ \citenamefont
  {Pfleiderer}}]{krey2012first}%
  \BibitemOpen
  \bibfield  {author} {\bibinfo {author} {\bibfnamefont {C.}~\bibnamefont
  {Krey}}, \bibinfo {author} {\bibfnamefont {S.}~\bibnamefont {Legl}}, \bibinfo
  {author} {\bibfnamefont {S.~R.}\ \bibnamefont {Dunsiger}}, \bibinfo {author}
  {\bibfnamefont {M.}~\bibnamefont {Meven}}, \bibinfo {author} {\bibfnamefont
  {J.~S.}\ \bibnamefont {Gardner}}, \bibinfo {author} {\bibfnamefont {J.~M.}\
  \bibnamefont {Roper}}, \ and\ \bibinfo {author} {\bibfnamefont
  {C.}~\bibnamefont {Pfleiderer}},\ }\href@noop {} {\bibfield  {journal}
  {\bibinfo  {journal} {Physical review letters}\ }\textbf {\bibinfo {volume}
  {108}},\ \bibinfo {pages} {257204} (\bibinfo {year} {2012})}\BibitemShut
  {NoStop}%
\bibitem [{\citenamefont {Samarakoon}\ \emph {et~al.}(2022)\citenamefont
  {Samarakoon}, \citenamefont {Sokolowski}, \citenamefont {Klemke},
  \citenamefont {Feyerherm}, \citenamefont {Meissner}, \citenamefont {Borzi},
  \citenamefont {Ye}, \citenamefont {Zhang}, \citenamefont {Dun}, \citenamefont
  {Zhou} \emph {et~al.}}]{samarakoon2022structural}%
  \BibitemOpen
  \bibfield  {author} {\bibinfo {author} {\bibfnamefont {A.~M.}\ \bibnamefont
  {Samarakoon}}, \bibinfo {author} {\bibfnamefont {A.}~\bibnamefont
  {Sokolowski}}, \bibinfo {author} {\bibfnamefont {B.}~\bibnamefont {Klemke}},
  \bibinfo {author} {\bibfnamefont {R.}~\bibnamefont {Feyerherm}}, \bibinfo
  {author} {\bibfnamefont {M.}~\bibnamefont {Meissner}}, \bibinfo {author}
  {\bibfnamefont {R.~A.}\ \bibnamefont {Borzi}}, \bibinfo {author}
  {\bibfnamefont {F.}~\bibnamefont {Ye}}, \bibinfo {author} {\bibfnamefont
  {Q.}~\bibnamefont {Zhang}}, \bibinfo {author} {\bibfnamefont
  {Z.}~\bibnamefont {Dun}}, \bibinfo {author} {\bibfnamefont {H.}~\bibnamefont
  {Zhou}},  \emph {et~al.},\ }\href@noop {} {\bibfield  {journal} {\bibinfo
  {journal} {Physical Review Research}\ }\textbf {\bibinfo {volume} {4}},\
  \bibinfo {pages} {033159} (\bibinfo {year} {2022})}\BibitemShut {NoStop}%
\bibitem [{\citenamefont {St{\"o}ter}(2019)}]{stoter2019static}%
  \BibitemOpen
  \bibfield  {author} {\bibinfo {author} {\bibfnamefont {T.}~\bibnamefont
  {St{\"o}ter}},\ }\emph {\bibinfo {title} {Static and dynamic magnetoelastic
  properties of spin ice}},\ \href@noop {} {Ph.D. thesis},\ \bibinfo  {school}
  {Dissertation, Dresden, Technische Universit{\"a}t Dresden, 2019} (\bibinfo
  {year} {2019})\BibitemShut {NoStop}%
\bibitem [{\citenamefont {Ma}\ \emph {et~al.}(2015)\citenamefont {Ma},
  \citenamefont {Cui}, \citenamefont {Ueda}, \citenamefont {Tang},
  \citenamefont {Chen}, \citenamefont {Tamura}, \citenamefont {Wu},
  \citenamefont {Fujioka}, \citenamefont {Tokura},\ and\ \citenamefont
  {Shen}}]{ma2015mobile}%
  \BibitemOpen
  \bibfield  {author} {\bibinfo {author} {\bibfnamefont {E.~Y.}\ \bibnamefont
  {Ma}}, \bibinfo {author} {\bibfnamefont {Y.-T.}\ \bibnamefont {Cui}},
  \bibinfo {author} {\bibfnamefont {K.}~\bibnamefont {Ueda}}, \bibinfo {author}
  {\bibfnamefont {S.}~\bibnamefont {Tang}}, \bibinfo {author} {\bibfnamefont
  {K.}~\bibnamefont {Chen}}, \bibinfo {author} {\bibfnamefont {N.}~\bibnamefont
  {Tamura}}, \bibinfo {author} {\bibfnamefont {P.~M.}\ \bibnamefont {Wu}},
  \bibinfo {author} {\bibfnamefont {J.}~\bibnamefont {Fujioka}}, \bibinfo
  {author} {\bibfnamefont {Y.}~\bibnamefont {Tokura}}, \ and\ \bibinfo {author}
  {\bibfnamefont {Z.-X.}\ \bibnamefont {Shen}},\ }\href@noop {} {\bibfield
  {journal} {\bibinfo  {journal} {Science}\ }\textbf {\bibinfo {volume}
  {350}},\ \bibinfo {pages} {538} (\bibinfo {year} {2015})}\BibitemShut
  {NoStop}%
\bibitem [{\citenamefont {Chalker}\ \emph {et~al.}(2011)\citenamefont
  {Chalker}, \citenamefont {Lacroix}, \citenamefont {Mendels},\ and\
  \citenamefont {Mila}}]{chalker2011introduction}%
  \BibitemOpen
  \bibfield  {author} {\bibinfo {author} {\bibfnamefont {J.~T.}\ \bibnamefont
  {Chalker}}, \bibinfo {author} {\bibfnamefont {C.}~\bibnamefont {Lacroix}},
  \bibinfo {author} {\bibfnamefont {P.}~\bibnamefont {Mendels}}, \ and\
  \bibinfo {author} {\bibfnamefont {F.}~\bibnamefont {Mila}},\ }\href@noop {}
  {\emph {\bibinfo {title} {Introduction to Frustrated Magnetism: Materials,
  Experiments, Theory}}}\ (\bibinfo  {publisher} {Springer-Verlag Berlin
  Heidelberg},\ \bibinfo {year} {2011})\BibitemShut {NoStop}%
\end{thebibliography}%

\end{document}